\documentclass[onecolumn,showpacs,aps,epsfig,nofootinbib]{revtex4}

\usepackage{graphicx}
\usepackage{epstopdf}
\usepackage{latexsym}
\usepackage{graphicx,times}
\usepackage{natbib}
\usepackage{amssymb,amsmath}
\usepackage{color}
\usepackage{colordvi}

\usepackage[center]{subfigure}

\begin{document}

 \newcommand{\bq}{\begin{equation}}
 \newcommand{\eq}{\end{equation}}
 \newcommand{\bqn}{\begin{eqnarray}}
 \newcommand{\eqn}{\end{eqnarray}}
 \newcommand{\nb}{\nonumber}
 \newcommand{\lb}{\label}
 \newcommand{\tc}{\textcolor{black}}
\newcommand{\PRL}{Phys. Rev. Lett.}
\newcommand{\PL}{Phys. Lett.}
\newcommand{\PR}{Phys. Rev.}
\newcommand{\CQG}{Class. Quantum Grav.}


%

\title{Probing the statistical properties of CMB $B$-mode polarization through Minkowski Functionals}

\author{Larissa Santos}
\author{Kai Wang}
\author{Wen Zhao}
\email[]{wzhao7@ustc.edu.cn}
\affiliation{CAS Key Laboratory for Researches in Galaxies and Cosmology, Department of Astronomy, University of Science and Technology of China, Chinese Academy of Sciences, Hefei, Anhui 230026, China}

\date{\today}

\begin{abstract}
The detection of the magnetic type $B$-mode polarization is the main goal of future cosmic microwave background (CMB) experiments. In the standard model, the $B$-mode map is a strong non-gaussian field due to the CMB lensing component. Besides the two-point correlation function, the other statistics are also very important to dig the information of the polarization map. In this paper, we employ the Minkowski functionals to study the morphological properties of the lensed $B$-mode maps. We find that the deviations from Gaussianity are very significant for both full and partial-sky surveys. As an application of the analysis, we investigate the morphological imprints of the foreground residuals in the $B$-mode map. We find that even for very tiny foreground residuals, the effects on the map can be detected by the Minkowski functional analysis. Therefore, it provides a complementary way to investigate the foreground contaminations in the CMB studies.


\end{abstract}


\maketitle

\section{ Introduction}

The temperature and polarization anisotropies of the cosmic microwave background (CMB) radiation contain useful cosmological information (see for instance, \cite{cmb-books,challinor1,challinor2}), playing a crucial role in constraining the cosmological parameters and testing the cosmological principle in modern cosmology \cite{wmap9,planck2015}. In the standard model, the CMB is a linearly polarized photon field, which is completely described by the stocks parameters $T(\hat{\gamma})$, $Q(\hat{\gamma})$ and $U(\hat{\gamma})$. $T$ is a two-dimensional random scalar field, which describes the CMB temperature anisotropy. $Q$ and $U$, which are not scalar fields in mathematics, describe the linear polarization.  It is then convenient to define the electric-type (i.e. $E$-mode) and magnetic-type (i.e. $B$-mode) polarization from the observables $Q$ and $U$ \cite{zaldarriaga-b-mode,kamionkowski-b-mode}. So, equivalently, $T$, $E$ and $B$ maps compose all the CMB information.


Nowadays, due to very precise observations from WMAP and Planck satellites, the detection of $T$ map is close to the cosmic variance limit \cite{wmap9,planck2015}. For $E$-mode polarization, the recent released Planck data have also shown very precise observations, and quite close to the cosmic variance limit for low multipoles $\ell\lesssim 1000$ \cite{planck-e-mode}. Therefore, the detection of $B$-mode polarization becomes the main goal of future CMB experiments \cite{cmb-taskforce}. In the standard model, the $B$-mode polarization is generated by two sources:  The primordial gravitational waves (i.e. tensor perturbations) \cite{grishchuk,inflation}, which are important in the low-multipole range \cite{zaldarriaga-b-mode,kamionkowski-b-mode,gw-b-mode}, and the cosmic weak lensing \cite{zaldarriaga-lensing,hu-lensing,lewis-review}, which is dominant at high multipoles. The $B$-mode observable is a combination of these two components. In the past two years, several ground-based experiments, including SPTPol \cite{sptpol}, POLARBEAR \cite{polarbear}, ACTPol \cite{actpol}, BICEP2 and Keck Array \cite{bicep2-1,bicep2-2,bicep2-3,bicep2-4}, as well as Planck satellite \cite{planck-b-mode} have detected the definite lensed $B$-mode signal in the high-multipole range. It is also expected that Planck detects the $B$-mode signal in the low multipoles for the first time in the forthcoming year \cite{planck-white}. For the potential observations in the near future, various ground-based and balloon-borne CMB experiments, including BICEP3, AdvACT, CLASS, Simons Array, SPT-3G, C-BASS, QUIJOTE, EBEX, QUBIC, QUIET, PIPER, Spider, LSPE, et al. \cite{ground-based}, are expected to have much smaller instrumental noise and also multi-frequency channels,
enabling the observation of the $B$-mode polarization at a very high confidence level. In addition, the next satellite generation of CMB experiments, including LiteBIRD \cite{litebird}, CMBPOL \cite{cmbpol}, COrE \cite{core}, PRISM \cite{prism}, PIXIE \cite{pixie} et al., will provide an excellent opportunity to precisely observe the $B$-mode in the full multipole range.

As the $B$-mode polarization era approaches (see for instance \cite{efstathiou2010,krauss2010,kamionkowski2015}), careful studies on the properties of the $B$-mode field are necessary. In previous works, most of the efforts have focused on the second-order power spectrum $C_{\ell}^{BB}$, which is equivalent to the two-point correlation function of the $B$-mode map.
The two-point correlation function can completely describe the statistical properties of a purely Gaussian field \cite{probability-book,probability-book2}. {However, it is well known that due to the lensing effect, the total two-dimensional $B$-mode map is a non-Gaussian field. \footnote{Several methods have been developed to reconstruct the lensing field and delense the $B$-mode map by using the CMB data alone \cite{hu-delense,other-delense} or by combining the CMB data with external data \cite{delense2,cmbpol-lense}. In this paper, we will not consider the effect of delensing.}} In addition to the power spectrum, other statistics (i.e. the Minkowski functionals, higher-order correlators, and so on) are also important in the case of non-gaussian field for two reasons. First, these statistics contain extra cosmological information, so comparing with the power spectrum, they might provide complementary channels to constrain the cosmological parameters (such as the neutrinos). Second, these statistics may be helpful to distinguish cosmological $B$-mode component and contaminated $B$-mode component (i.e. the residuals of foreground radiations \cite{foregrounds,planck-foregrounds}, the residuals of the $E$-$B$ mixture \cite{ebmixture1,ebmixture2,ebmixture3}, and so on), according to their different statistical properties. In this paper, as a first step, we shall apply the Minkowski functionals (MFs) to study the morphological characteristics of the lensed $B$-mode polarization. We shall focus on how it deviates from Gaussianity. We find that, even in the extreme case in which only a small part of the sky is surveyed, instrumental noise is added and gravitational waves are considered, the $B$-mode map has the definite non-gaussian distribution. It deviates from Gaussianity at very high confidence level. As an application of the MF analysis, we also study the imprints of the foreground residuals in the $B$-map. We find that for the nearly full sky survey, the deviation of MFs caused by the residuals can be detected if more than $0.4\%$ of foreground radiation is left as residuals in the $B$-map. For surveys with a small sky coverage, the deviation can be detected if more than $1\%$ of residuals are left. For both cases, the contributions of residuals on the CMB power spectrum are very small, so the MFs provide a new and powerful tool to study the different components in the observable $B$-maps.

The outline of the paper is as follows. In Sec. 2, we briefly introduce the basic characteristics of the CMB polarization. In Sec. 3, we introduce the MFs used in our analysis. In Sec. 4, by employing the MFs, we study the morphologic properties of $B$-mode polarization, and study the deviations from Gaussianity. In Sec. 5, we focus on the morphological imprints of the foreground residuals on the $B$-map. Finally, Sec. 6 is contributed as a conclusion of this article.

\section{$E$- and $B$-mode polarization of CMB}

As well known, the CMB is a linear polarized radiation field in the two-dimensional sphere. In addition to the temperature anisotropy $\Delta T(\hat{\gamma})$ field, the observable quantities for linear polarization include the Stocks parameters: $Q(\hat{\gamma})$ and $U(\hat{\gamma})$. Since they are not scalar fields, it is necessary to construct the so-called $E$- and $B$-mode polarization fields for the data analysis. For mathematical simplicity, it is convenient to introduce the complex conjugate polarization fields $P_{\pm}$ as follows:
 $P_{\pm}(\hat{\gamma})\equiv Q(\hat{\gamma}) \pm i U(\hat{\gamma})$,
where $\hat{\gamma}$ denotes the position, and $Q$ and $U$ are assumed to be real fields in the sky. The fields $P_{\pm}$, being $\pm2$ spin-weighted quantities, can be expanded over appropriate spin-harmonic functions \cite{seljak1996}
 $P_{\pm}(\hat{\gamma})=\sum_{\ell m} a_{\pm2,\ell m}~_{\pm 2}Y_{\ell m}(\hat{\gamma})$,
where $_{\pm 2}Y_{\ell m}(\hat{\gamma})$ are the spin-weighted spherical harmonics. The $E$- and $B$-mode multipoles are defined in terms of the coefficients $a_{\pm 2,\ell m}$ in the following manner:
 $E_{\ell m}\equiv -\frac{1}{2}[a_{2,\ell m}+a_{-2,\ell m}]$,
 $B_{\ell m}\equiv -\frac{1}{2i}[a_{2,\ell m}-a_{-2,\ell m}]$.
One can now define the electric polarization sky map $E(\hat{\gamma})$ and the magnetic polarization sky map $B(\hat{\gamma})$ as
 \begin{equation}
 E(\hat{\gamma})\equiv \sum_{\ell m}E_{\ell m} Y_{\ell m}(\hat{\gamma}),~~
 B(\hat{\gamma})\equiv \sum_{\ell m}B_{\ell m} Y_{\ell m}(\hat{\gamma}).
 \end{equation}
It can be proved that the constructed $E(\hat{\gamma})$ is a scalar field, and $B(\hat{\gamma})$ is a pseudo-scalar field. The power spectra of the $E$ and $B$ modes are defined as in general,
 \begin{eqnarray}
 C_{\ell}^{EE}&\equiv&\frac{1}{2\ell+1}\sum_m\langle E_{\ell m} E_{\ell m}^* \rangle, \\
 C_{\ell}^{BB}&\equiv&\frac{1}{2\ell+1}\sum_m\langle B_{\ell m} B_{\ell m}^* \rangle,
 \end{eqnarray}
where the brackets denote the average over all realizations. For the purely Gaussian field, all the statistical properties are encoded in the second-order power spectrum.

In the standard inflation+$\Lambda$CDM cosmology, the $B$-mode polarization is generated by two sources. One is the primordial gravitational waves (i.e. tensor perturbations) $h_{ij}(\vec{k})$, which satisfy the Gaussian distribution in the inflationary scenario. In the linear approximation, the $B$-mode coefficients, $B_{\ell m}$, linearly depend on the perturbations $h_{ij}(\vec{k})$ \cite{gw-b-mode}. The generated $B$-mode map $B(\hat{\gamma})$ is then a random Gaussian field. The other source of $B$-mode polarization is the weak gravitational lensing of CMB photons due to large-scale structure along the line of sight, which results in a fractional conversion of the $E$-mode (from density perturbations) to the $B$-mode component \cite{zaldarriaga-lensing,hu-lensing,lewis-review,knox-lensing}. The lensed $B$-mode polarization directly relates to the large-scale structure in the late Universe, and its distribution is definitely non-gaussian. In general, the observed $B$-mode map is the combination of these two parts, in which the component generated by gravitational waves always dominates in the large angular scales, i.e. low multipoles, and the component caused by weak lensing is dominant in the small angular scales. Therefore, the total $B$-mode map satisfies the non-gaussian distribution. For the two-dimensional non-gaussian field, in addition to the second-order power spectrum, the other non-gaussian statistics (such as the angular bispectrum and trispectrum, other higher-order statistics, MFs, clustering strength, Betti numbers, and so on) are important to study the statistical properties of the field. In this paper, we shall analyze the statistical properties, especially deviations from Gaussianity of the CMB $B$-mode polarization map by employing the MFs.

\section{Minkowski Functionals}

Minkowski functionals characterize the morphological properties of convex, compact sets in an $n$-dimensional space \cite{minkowski1903,gott1990,schneider1993,mecke1994,schmalzing1997,schmalzing1998,wizitzki1998}. On the two-dimensional spherical surface $\mathcal{S}^2$, any morphological property can be expanded as a linear combination of three MFs, which represent the area, contour length, and integrated geodetic curvature of an excursion set \cite{gott1990,schmalzing1998}. For a given threshold $\nu$, it is convenient to define the excursion set $Q_{\nu}$ and its boundary $\partial Q_{\nu}$ of a smooth scalar field $u$ as follows: $Q_{\nu}=\{x\in \mathcal{S}^2|u(x)>\nu\}$ and $\partial Q_{\nu}=\{x\in \mathcal{S}^2|u(x)=\nu\}$. Then, the MFs $v_0$, $v_1$ and $v_2$ can be written as \cite{schmalzing1998}
 \begin{equation}
 v_0(\nu)=\int_{Q_{\nu}}\frac{da}{4\pi},~v_1=\int_{\partial Q_{\nu}}\frac{d l}{16\pi},~v_2=\int_{\partial Q_{\nu}}\frac{\kappa dl}{8\pi^2},
 \end{equation}
where $da$ and $dl$ denote the surface element of $\mathcal{S}^2$ and the line element along $\partial Q_{\nu}$, respectively, and $\kappa$ is the geodetic curvature. { Given a pixelized map $u(x_i)$ with equal-area pixelization, the first MF $\nu_0$ can be numerically calculated as \cite{schmalzing1998,lim2012}}:
 \begin{equation}
 v_0(\nu)=\frac{1}{N_{\rm pix}} \sum_{k=1}^{N_{\rm pix}} \Theta(u-\nu),\label{eq2}
 \end{equation}
{where $\Theta$ is the Heaviside step function.} The other two MFs $\nu_1$ and $\nu_2$ can be calculated by the following formulae \cite{schmalzing1998,lim2012},
 \begin{equation}
 v_i(\nu)=\frac{1}{N_{\rm pix}} \sum_{k=1}^{N_{\rm pix}} \mathcal{I}_i(\nu,x_k),~~(i=1,2),\label{eq2}
 \end{equation}
where
 \begin{eqnarray}
 \mathcal{I}_1(\nu,x_k)&=&\frac{\delta(u-\nu)}{4}\sqrt{u_{;\theta}^2+u_{;\phi}^2}, ~~~
 \mathcal{I}_2(\nu,x_k) = \frac{\delta(u-\nu)}{2\pi}\frac{2u_{;\theta}u_{;\phi}u_{;\theta\phi}-u_{;\theta}^2u_{;\phi\phi}-u_{;\phi}^2u_{;\theta\theta}}{u_{;\theta}^2+u_{;\phi}^2}.
 \end{eqnarray}
Note that $u_{;i}$ denotes the covariant differentiation of $u$
with respect to the coordinate $i$. However, in the numerical computations, the delta function in these
formulae can be numerically approximated through a discretization
of threshold space in bins of width $\Delta\nu$ by the
{{Heaviside step function}}
$\delta_N(x)=(\Delta\nu)^{-1}[\Theta(x+\Delta\nu/2)-\Theta(x-\Delta\nu/2)]$.

The expectation values of three MFs for a Gaussian random field
are derived in \cite{tomita1986}, which have been explicitly
expressed in \cite{schmalzing1998,lim2012}. For $v_0$, the expectation value for a Gaussian random field $u_g$ is given by
 \begin{equation}
 \langle{v}_0^g(\nu)\rangle=\frac{1}{2}\left(1-{\rm erf}\left(\frac{\nu-\mu}{\sqrt{2}\sigma}\right)\right),
 \label{v0_bar}
 \end{equation}
where erf is the Gaussian error function. The mean and variance of $u$, and the variance of its gradient are give by
 \begin{equation}
 \mu\equiv\langle u\rangle,~\sigma^2\equiv\langle u^2\rangle-\mu^2,~\tau\equiv\frac{1}{2}\langle|\nabla u|^2\rangle.
 \end{equation}
The variance $\sigma^2$ and the variance of the gradient $\tau$ can also be expressed through the Gaussian angular power spectrum $C_{\ell}$ by
 \begin{equation}
 \sigma^2=\frac{1}{4\pi}\sum_{\ell}(2\ell+1)C_{\ell},
 ~~\tau=\frac{1}{4\pi}\sum_{\ell}(2\ell+1)C_{\ell}\frac{\ell(\ell+1)}{2}.
 \end{equation}
The expectation values for $v_1^g$ and $v_2^g$ can be approximated by the following formulae \cite{lim2012},
 \begin{eqnarray}
 \langle{v}_{1}^{g}(\nu)\rangle&=&\frac{1}{8}\frac{\sqrt{\tau}}{\Delta\nu}\sqrt{\frac{\pi}{2}}\left[{\rm erf}\left(\frac{\nu-\mu+\Delta\nu/2}{\sqrt{2}\sigma}\right)
 -{\rm erf}\left(\frac{\nu-\mu-\Delta\nu/2}{\sqrt{2}\sigma}\right)\right],\label{v1_bar}\\
  \langle{v}_{2}^{g}(\nu)\rangle&=&\frac{1}{(2\pi)^{3/2}}\frac{\tau}{\sigma^2}\frac{{\sigma}}{\Delta\nu}\left[{\rm exp}\left(-\frac{(\nu-\mu-\Delta\nu/2)^2}{2\sigma^2}\right)
  -{\rm exp}\left(-\frac{(\nu-\mu+\Delta\nu/2)^2}{2\sigma^2}\right)\right].
 \label{v2_bar}
 \end{eqnarray}

For some special cases of weak non-Gaussian CMB fields, the analytical formulae for the MFs are also derived in \cite{matsubara2003,hikage2006,matsubara2010}. These MFs have been applied by cosmologists to search for deviations from Gaussianity  in the CMB $T$-map and $E$-map, where the non-gaussianities are expected to be weak (see \cite{schmalzing1998,novikov1999,novikov2004,hikage2008,natoli2009,wmap-mink,lim2012,ducout2012,park2012,masato2012,hikage2012,munshi2012,fang2013,modest2012,zhao2014,park2014,planck-nongaussian} for instance). In this paper, we shall investigate the statistical properties of the CMB $B$-mode polarization field, in which the non-Gaussianity is expected to be quite large due to the contribution of the cosmic weak lensing. We will employ the MFs to quantify the deviation from Gaussianity of the CMB $B$-mode map, and study the possible imprints of various foreground contaminations.

\section{Non-gaussianities of $B$-mode polarization}

\subsection{The full-sky ideal case without noise}

 As the first step, we consider the ideal case, in which the $B$-mode polarization is only generated by gravitational waves and cosmic weak lensing. We assume the full-sky survey and ignore contamination from instrumental noise, foreground emissions and other effects. Throughout this paper, we adopt the standard cosmological model with the parameters derived from Planck 2015 results \cite{planck-parameter}, $\Omega_bh^2=0.02230,~\Omega_ch^2=0.1188,~100\theta_{\rm MC}=1.04093,~~\tau_{\rm rei}=0.066,~\ln(10^{10}A_s)=3.064,~n_s=0.9667$.
For the contribution of primordial gravitational waves, we consider four models in our discussion, in which the tensor-to-scalar ratio is $r=0$, $0.01$, $0.05$ and $0.1$, respectively. In all these models, we set the spectral index of gravitational waves as $n_t=0$.


{Throughout this paper, we use the public LensPix code \cite{lenspix} to simulate the lensed $a_{\ell m}^{TT}$, $a_{\ell m}^{EE}$ (i.e $E_{\ell m}$), $a_{\ell m}^{BB}$ (i.e. $B_{\ell m}$) coefficients and the corresponding lensed $T$, $Q$ and $U$ sky maps.
Under the weak lensing approximation, this package implements a pixel remapping approach to mimic the effects of lensing on the CMB photons \cite{lenspix-paper}. The numerical accuracy of LensPix has been widely checked: Regarding the CMB power spectra, the authors of LensPix showed that the lensed $C_{\ell}$ derived from the lensed sky simulations agree with the lensed power spectra derived from the CAMB package, which we also checked (see Fig. \ref{fig1}). {Moreover, \cite{smith2012} also found that the correlations of CMB power spectra numerical results coincide with the analytical ones (Note that in this paper, the authors adopted a different code, in which the algorithm is similar to the LensPix). Even so, in order to overcome the inaccuracy of LensPix at the high multipoles \cite{lenspix,fabbian2013}, throughout the calculations in this paper, we simulate multipoles up to $\ell_{\max}=1024$, but use only the multipole range of $\ell\le1000$ for analysis.} We adopt the resolution parameter $N_{\rm side}=512$ in this paper. For high accuracy, we adopt the cubic interpolation method with the interpolation factor of $1.5$. In addition, we apply the Gaussian smoothing with the parameter FWHM=$10'$ to smooth the high multipoles, where FWHM is the full width at a half maximum beam.}


In our calculation, for each case, we use the public CAMB code \cite{camb} to calculate the unlensed CMB power spectra $C_{\ell}^{XY}$ ($XY=TT,EE, BB,TE$). Then, we use the public LensPix code \cite{lenspix} to independently generate 1000 lensed CMB realizations, including the lensed $T$, $Q$ and $U$ maps ($E$- and $B$-mode maps are directly constructed from $Q$ and $U$ maps by the standard process introduced in Sec. II). The lensed power spectra $C_{\ell}^{XY}$ can be directly calculated by using these lensed maps.
In Fig. \ref{fig1}, we plot the CMB power spectra in both unlensed and lensed cases in the model with $r=0$. { The unlensed power spectra (red dashed lines in Fig. \ref{fig1}) are calculated using CAMB package. For the lensed power spectra, we have considered two cases: In the first one (blue solid lines), they are calculated using the CAMB package, and, in the second case (black solid lines),  we calculated the mean value of 1000 lensed samples, where the LensPix package is used to generate the simulated power spectra. Fig. \ref{fig1} shows that the spectra in these two lensed cases are consistent with each other. {For instance, for the $TT$ spectrum at $\ell=200$, we get $\ell(\ell+1)C_{\ell}^{TT}/2\pi=5458.3\mu$K$^2$ and $5447.6\mu$K$^2$ in these two cases, respectively. The difference between them is smaller than $0.2\%$. For the $BB$ spectra, the difference is slightly larger: For $\ell=200$, we get $\ell(\ell+1)C_{\ell}^{BB}/2\pi=0.0112\mu$K$^2$ and $0.0109\mu$K$^2$ in these two cases, respectively. While for $\ell=900$, we have $\ell(\ell+1)C_{\ell}^{BB}/2\pi=0.00780\mu$K$^2$ and $0.00745\mu$K$^2$ in these two cases, respectively. The difference in both multipoles are smaller than $5\%$.} Comparing the unlensed spectra with the lensed ones, we find that the $TT$, $TE$ and $EE$ power spectra are smoothed by the lensing effect, which is consistent with the physical predictions \cite{lewis-review}. In this case, the $BB$ power spectrum is only sourced by the lensing effect, so it vanishes for the unlensed CMB power spectrum. Due to the weak lensing effect, all these three maps include the non-gaussian signal. The $B$-mode is dominated by the contribution of weak lensing in small scales even if we include the contribution of gravitational waves, so we expect the non-Gaussianity in the $B$-map to be much larger than that in the other two maps.

\begin{figure}[t]
\centerline{\includegraphics[width=16cm,height=12cm]{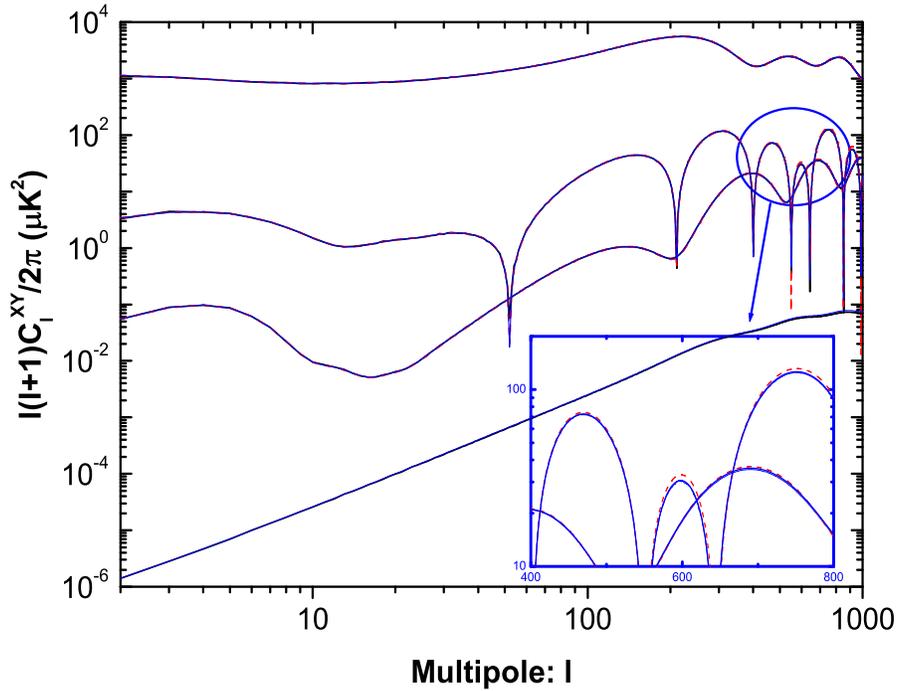}}
\caption{The theoretical lensed (blue solid lines) and unlensed (red dashed lines) CMB power spectra $C_{\ell}^{XY}$ calculated by using CAMB package. The lines from upper to lower denote the results with $XY=TT$, $TE$, $EE$ and $BB$ respectively. For $XY=TE$, the figure shows the absolute value of the spectrum. In this figure, we also plot the simulated lensed power spectra $C_{\ell}^{XY}$ (black solid lines), which are {the mean value of 1000 simulated lensed samples. In this lensing simulations, we have used LensPix package. We find that the simulated lensed power spectra overlap with the theoretical lensed ones, which shows that the lensing simulations are quite accurate.} Note that, we have adopted the model with $r=0$ in this figure.}
\label{fig1}
\end{figure}

{In general, the measurement of MFs requires smoothing of CMB maps in order to remove the contribution of high multipoles (which are always dominated by the noises). The MFs derived from different smoothing scales always have significant correlation, since they are calculated based on the same CMB map. However, the information of CMB at different multipole range will be dominant if one chooses different smoothing scale, and the complementary effect between them are also very important, which have been widely confirmed by the previous works \cite{hikage2006,hikage2008,fang2013,ducout2012}. So, in the calculation, this smoothing procedure should be performed at various scales on a pixelized map in order to extract all the statistical information available \footnote{For the case with Gaussian or nearly Gaussian distributed field, in \cite{ducout2012} the authors introduced the Wiener filter and its first and second derivatives for the field. They found that the results obtained with a simple combination of Wiener filters set are similar (or even better) than the Gaussian smoothing at various scale. This method has been applied by Planck team to constrain the non-Gaussianity of CMB $T$ and $E$ maps \cite{planck-nongaussian}. However, in the present paper, we are working on the $B$-mode polarization map, which is expected to be highly non-gaussian, and the Wiener filter cannot be applied directly. So, in the calculation, we will adopt the traditional method by smoothing the maps at various scales, and combining them to extract all the statistical information.}. Adopting the methods developed in the previous works ($e.g.$ \cite{schmalzing1998,hikage2008,lim2012,zhao2014}), we calculate the MFs of the CMB maps as the following steps. For each realization (i.e. the simulated $B$-mode map with $N_{\rm side}=512$ and FWHM=$10'$), we smooth the map again using a Gaussian filter with a smoothing scale of $\theta_s$. In our calculation, the smoothing parameter assumes six different values, $\theta_s$=$10'$, $20'$, $30'$, $40'$, $50'$, $60'$, and we numerically calculate the MFs $v_i(\nu)$ for each smoothed map by the method introduced in Sec. III.} Note that the binning range of threshold $\nu$ is set to be from $-3.0$ to $3.0$ with 25 equally spaced bins of $\nu/\sigma$ ($\sigma$ is the standard deviation of the corresponding map) per each MF. The mean value and the error bar of $v_i(\nu)$ are numerically calculated by
\begin{equation}
\bar{v}_i(\nu)=\frac{1}{N}\sum_{k=1}^{N}v_i^{(k)}(\nu),~\Delta v_i(\nu)=\sqrt{\frac{1}{N-1}\sum_{k=1}^{N}\left[v_i^{(k)}(\nu)-\bar{v}_i(\nu)\right]^2},
\end{equation}
where $N=1000$ is the number of the samples, $v_i^{(k)}(\nu)$ is the $i$-th MF calculated from the $k$-th random sample. In Fig. \ref{fig2}, we plot the $\bar{v}_i(\nu)$ and the corresponding error bars for the case with $r=0$ and $\theta_s=10'$ (red curves in upper panels).

In this paper, we study the statistical properties of the lensed CMB maps by comparing them with the Gaussian cases. We investigate the deviation of the lensed maps from Gaussianity.
In our calculation, for a fair comparison, we consider the Gaussian maps which have the exact same power spectra $C_{\ell}^{XY}$ as the lensed maps. The analytical formulae for $\langle v_i^g(\nu)\rangle$ are given in Eqs. (\ref{v0_bar}), (\ref{v1_bar}) and (\ref{v2_bar}). In the upper panels of Fig. \ref{fig2}, we plot the analytical results as black solid lines for comparison. For the cross-checking, we also calculate the $v_i^g(\nu)$ by independently simulating 1000 Gaussian maps using the HEALPix package, where the input CMB power spectra $C_{\ell}^{XY}$ have exactly the same values of those in the lensed cases. Then, by repeating the same steps above, we calculate the MFs for the Gaussian samples. The mean values $\bar{v}^g_i(\nu)$ and the standard deviations $\Delta v^g_i(\nu)$ of Gaussian maps are also derived, which have also been plotted in the upper panels of Fig.\ref{fig2} as blue curves.

\begin{figure}[t]
\centerline{\includegraphics[width=18cm,height=8cm]{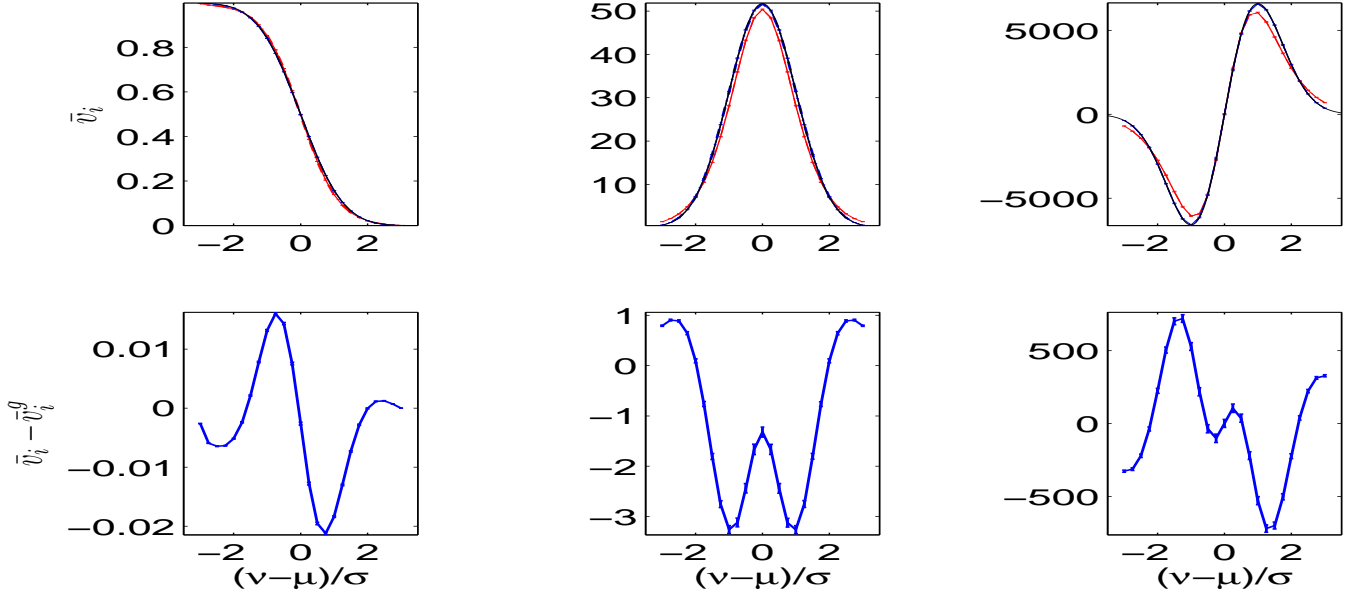}}
\caption{The upper panels denote the average MFs $\bar{v}_i$ and corresponding error bars $\Delta v_i$ for $B$-map. In each panel, the red line is the result for the lensed map, the black line is for the result of the Gaussian map, and blue line is for the analytical curve for the Gaussian case. The lower panels are the difference $\bar{v}_i-\bar{v}_i^g$. The left panels are for $i=0$, middle ones are for $i=1$ and right ones are for $i=2$. Note that, in this figure, we have adopted $r=0$ and the $\theta_s$=$10'$.}
\label{fig2}
\end{figure}

{As anticipated, Fig. \ref{fig2} shows that the results for the simulations and the analytical calculation, considering the Gaussian cases, are consistent with each other. {We have checked that the differences between them are less than $0.2\%$ for any MF, which clearly indicates the accuracy of our MF numerical code.} Since for both Gaussian maps and lensed maps we have used the exact same MF code, we are confident to state that our MF numerical calculations are accurate enough in all the cases considered here.} Obviously, we find that the MFs in the lensed $B$-mode polarization are significantly different from the Gaussian case for all the threshold values $\nu$. In order to show these differences more clearly, in Fig. \ref{fig2} (lower panels), we plot the values of $\delta\bar{v}_i(\nu)\equiv\bar{v}_i(\nu)-\bar{v}_i^g(\nu)$ for a given $\nu$, where we find that the value $\delta\bar{v}_i(\nu)$ is much larger than the statistical error for any given $\nu$. So, due to the weak lensing effect, the non-Gaussianity in the $B$-mode polarization is very large. \emph{For all three MFs, we find that the deviation from Gaussianity is mainly in the regions $\nu/\sigma\in(1,2)$ and $\nu/\sigma\in(-2,-1)$.  Left panels show that, in both regions, there are more pixels in the lensed maps than in the Gaussian ones. However, middle and right panels show that the total curve lengths and the total integrated geodetic curvatures in the lensed maps are relatively smaller than those in the Gaussian maps. These are the main morphological characters of the cosmic lensing effect in the CMB maps. These features seem to indicate that, comparing with the purely Gaussian distribution, there are more data in the lensed map that span between $1\sigma$ and $2\sigma$. They also indicate that the hot and cold spots have the tendency to form smooth clusters instead of the randomly distributed ones.} {In Appendix \ref{appendixB}, we applied the skewness and kurtosis statistics to probe the non-Gaussianity of the lensed $B$-maps. Although we found that these one-point statistics are less sensitive than the MFs, the distributions of skewness and kurtosis indicate similar lensing effects in the $B$-maps as the MFs.}

\begin{figure}[t]
\centerline{\includegraphics[width=18cm,height=18cm]{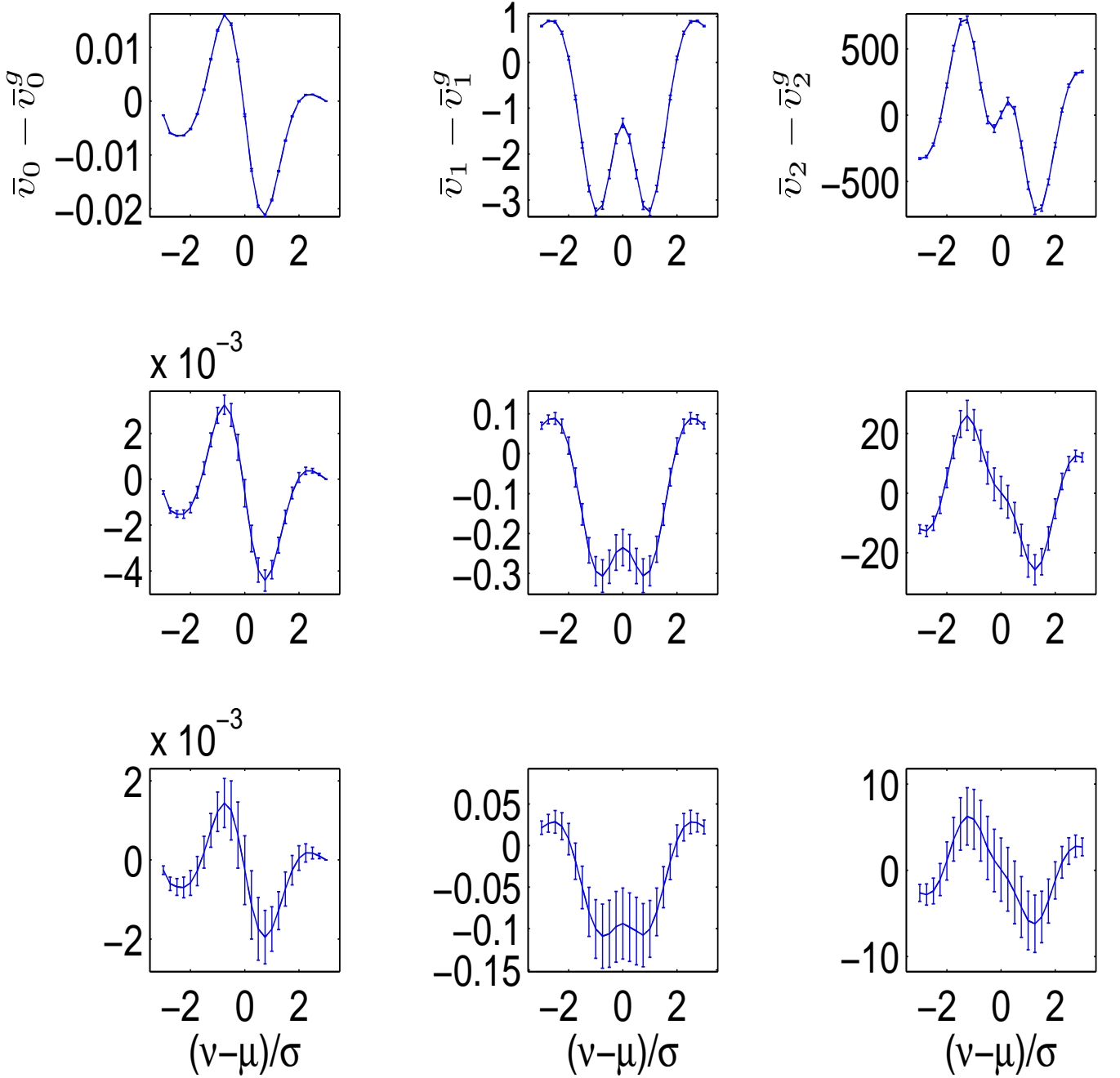}}
\caption{The difference between the MFs $v_i$ in the lensed $B$-mode map and the Gaussian $B$-mode map. The panels from upper and lower are the results for the cases with $\theta_s=10'$, $40'$, $60'$ respectively.
Note that, in this figure, we have adopted the model with $r=0$.}
\label{fig3}
\end{figure}

\begin{figure}[t]
\centerline{\includegraphics[width=18cm,height=18cm]{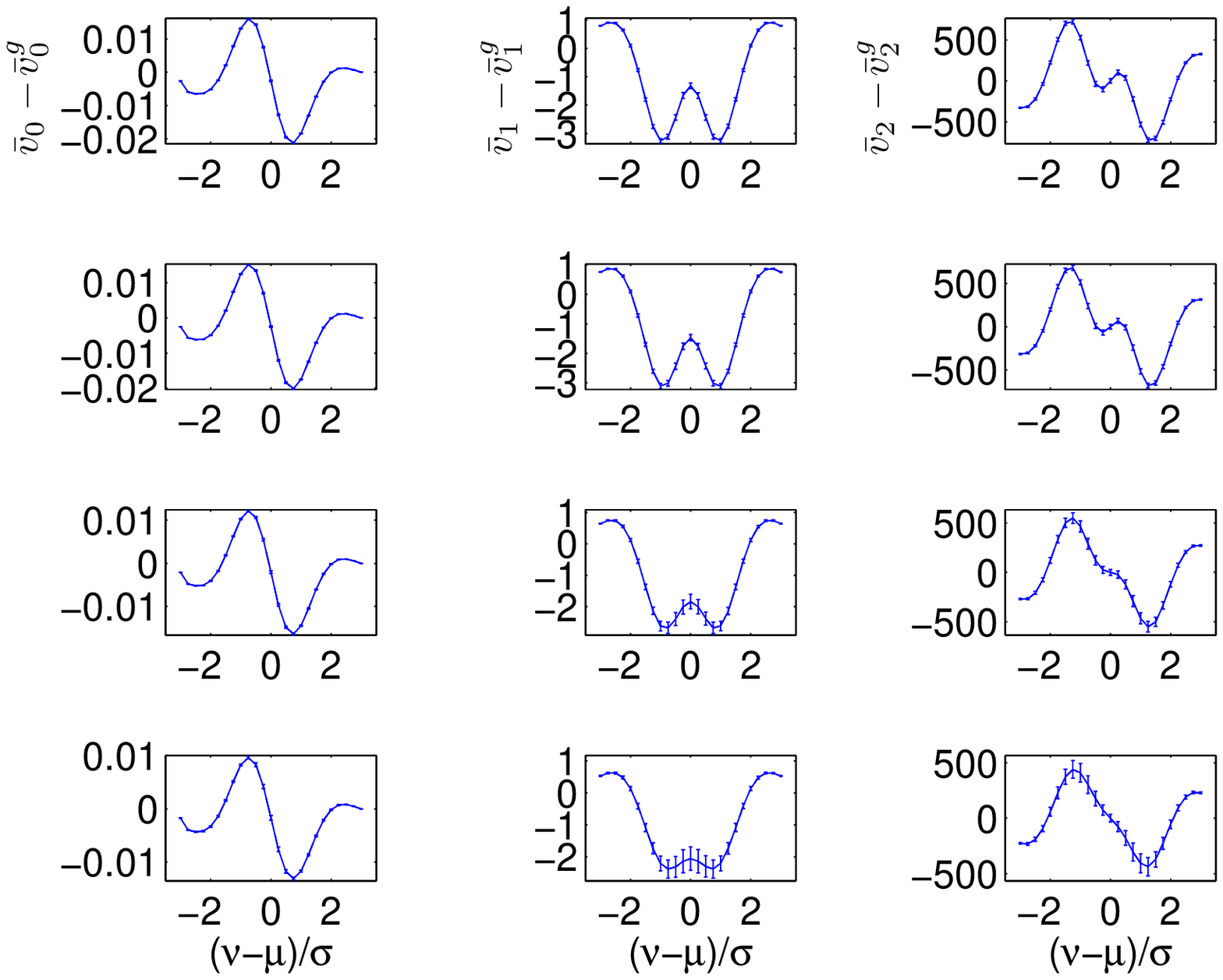}}
\caption{The difference between the MFs $v_i$ in the lensed $B$-mode map and the Gaussian $B$-mode map. The panels from upper and lower are the results for the models with $r=0$, $0.01$, $0.05$, $0.1$, respectively.
Note that, in this figure, we have adopted the smoothing parameter $\theta_s$=$10'$.}
\label{fig4}
\end{figure}

Of course, by increasing the gravitational waves contribution and/or smoothing scale parameter $\theta_s$, the non-Gaussianity becomes slightly smaller, which can be clearly found in Figs. \ref{fig3} and \ref{fig4}.
In both figures, we find that although the main morphological characters of lensing effect are still present for the cases with larger $r$ and/or larger $\theta_s$, the non-Gaussianity of the $B$-mode polarization becomes weaker. We quantify the significance of non-Gaussianity by defining the signal-to-noise ratio for the given smoothing scale parameter $\theta_s$ as
 \begin{equation}
 {\rm SNR}_{\theta_s} = \sum_{\alpha\alpha'}\left[\bar{v}_{\alpha}-\langle {v}^g_{\alpha}\rangle\right]\Sigma_{\alpha\alpha'}^{-1}
 \left[\bar{v}_{\alpha'}-\langle{v}^g_{\alpha'}\rangle\right],
 \end{equation}
where $\alpha$ and $\alpha'$ denote the binning number of threshold value $\nu$ and the different kinds of MF $i$, $\Sigma_{\alpha\alpha'}\equiv\frac{1}{N-1}\sum_{k=1}^{N}[({v}^{(k)}_{\alpha}-\bar{v}_{\alpha})({v}^{(k)}_{\alpha'}-{\bar{v}_{\alpha'}})]$ is the covariance matrix of $v_\alpha$. $\langle v_{\alpha}^{g}\rangle$ are the expectation values of MFs for the Gaussian random field, which are given by Eqs. (\ref{v0_bar}), (\ref{v1_bar}) and (\ref{v2_bar}). In all of our calculation, we use the average values of $v_{\alpha}^g$ derived from 1000 random Gaussian samples to replace the theoretical values. In Fig. \ref{fig5}, we plot the values of ${\rm SNR}_{\theta_s}$ for different $\theta_s$ values and different cosmological models, which clearly shows the dependence of non-Gaussianity of $B$-mode map on the smoothing scale. {Similar to the results for cases of CMB $T$ and $E$ maps (see \cite{ducout2012} for instance), we find that the value of ${\rm SNR}_{\theta_s}$ significantly decreases for larger $\theta_s$. So, most of signal is captured by the lowest $\theta_s$.}
In addition, from this figure, we also find that a larger gravitational-wave contribution follows a smaller non-Gaussianity in the map. Current observations constrain the amplitude of gravitational waves as $r\lesssim0.1$ \cite{planck-parameter}. But even for this model with $r=0.1$, the total $B$-mode polarization is significantly non-gaussian.

\begin{figure}[t]
\centerline{\includegraphics[width=12cm,height=10cm]{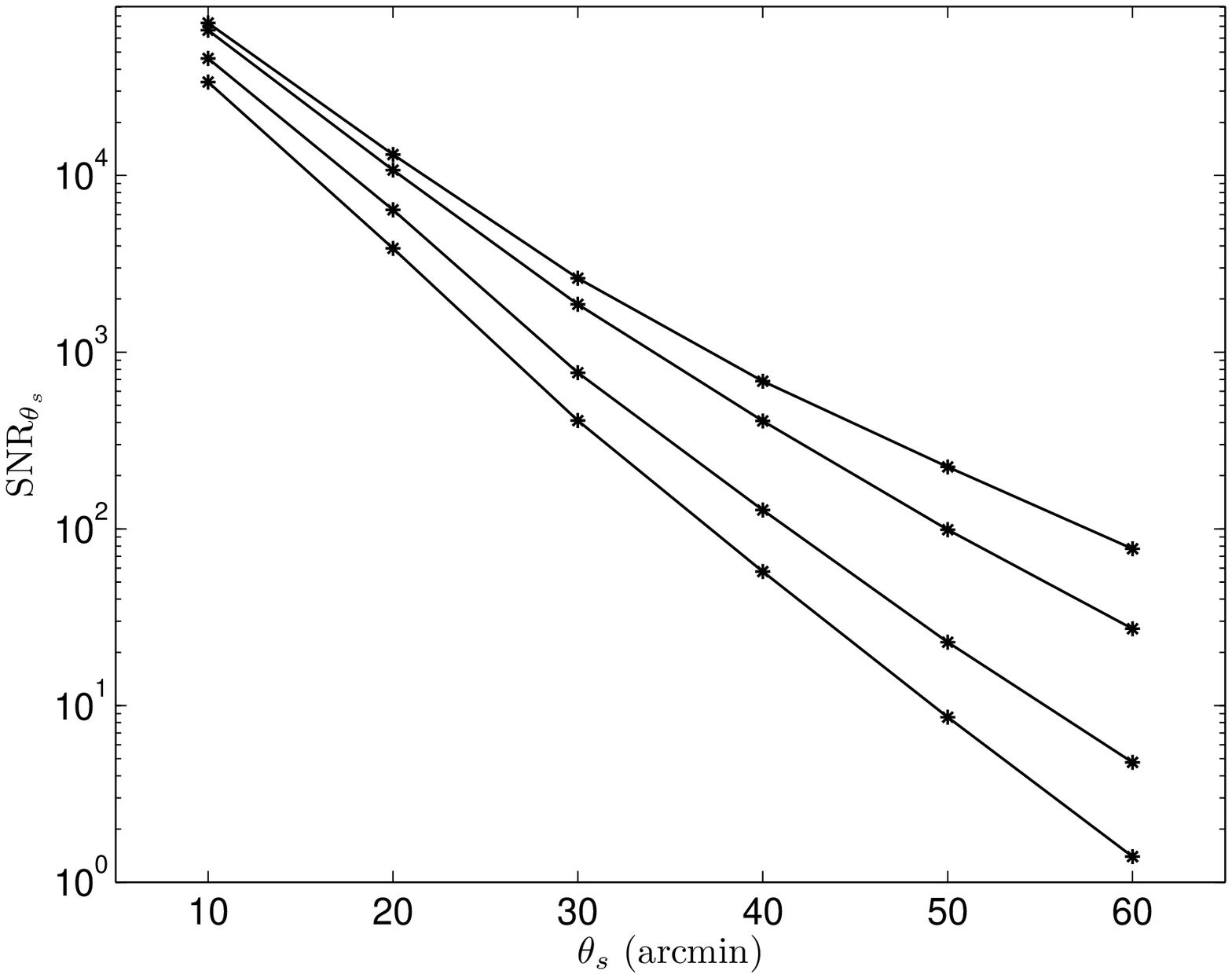}}
\caption{The SNR$_{\theta_s}$ depends on the smoothing parameter $\theta_s$ for the lensed $B$-map. From upper to lower, the lines denote the results in the models with $r=0$, $0.01$, $0.05$, $0.1$, respectively.}
\label{fig5}
\end{figure}

\begin{figure}[t]
\centerline{\includegraphics[width=16cm]{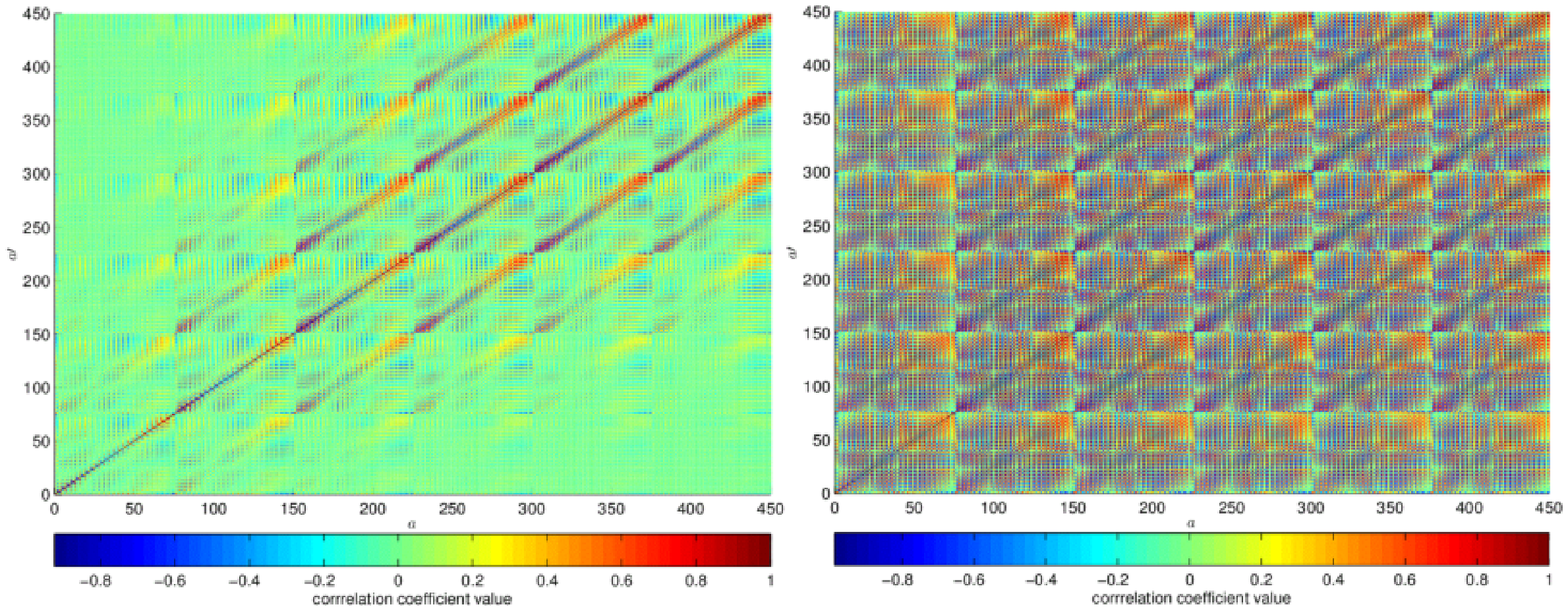}}
\caption{The correlation coefficients $\rho_{aa'}$ for the model with $r=0$ (left panel) and $r=0.1$ (right panel). Note that, in both panels, we have considered the ideal case without noises, sky-cut or foreground radiations. Note that, for each panel, $a(a')\le 75$ corresponds to MF with $\theta_s=10'$, $76<a(a')\le150$ corresponds to that with $\theta_s=20'$, and so on.}
\label{fig5.1}
\end{figure}

The total deviation from Gaussian distribution of the $B$-mode map can be quantified by the total signal-to-noise ratio as follows
 \begin{equation}
 {\rm SNR} = \sum_{aa'}\left[\bar{v}_{a}-\langle{v}^g_{a}\rangle\right]\Sigma_{aa'}^{-1}
 \left[\bar{v}_{a'}-\langle{v}^g_{a'}\rangle\right].
 \end{equation}
Note that here $a$ and $a'$ denote the binning number of threshold value $\nu$, the different kinds of MF $i$ and the smoothing scale parameter $\theta_s$. $\Sigma_{aa'}\equiv\frac{1}{N-1}\sum_{k=1}^{N}[({v}^{(k)}_{a}-\bar{v}_{a})({v}^{(k)}_{a'}-{\bar{v}_{a'}})]$. {As we have emphasized, the correlations between MFs with different smoothing scales $\theta_s$ are quite large, since they are obtained by the same $B$-mode map. So, the value of ${\rm SNR}$ is not a simple sum of different ${\rm SNR}_{\theta_s}$'s. In Fig. \ref{fig5.1}, we plot the correlation coefficient $\rho_{aa'}\equiv \Sigma_{aa'}/\sqrt{\Sigma_{aa}\Sigma_{a'a'}}$ for the model with $r=0$ (left panel) and $r=0.1$ (right panel), which clearly shows the strong correlation between different MFs.} By using the results above, we derive the values of ${\rm SNR}$ for different models, which are listed in Table \ref{table1}. Larger $r$ means larger contributions of primordial gravitational waves in the $B$-mode polarization, which satisfy the Gaussian distribution, and dilute the non-Gaussianity of the total $B$-map. For the extreme case without any gravitational waves contribution, i.e. $r=0$, the total SNR is $1.23\times10^5$, while in the opposite extreme case with $r=0.1$, the total SNR decreases to $0.64\times10^5$, reduced by a factor of 2.

\begin{table}[!htb]
\centering
\begin{tabular}{c @{\extracolsep{2em}} c c c c}
\hline\hline
 &  $r=0$  & $r=0.01$  &  $r=0.05$  &  $r=0.1$\\
\hline
T-map & 3.35 &  3.53 &  3.67 & 3.82 \\
E-map  & 7.07 & 6.77 &  6.89 & 6.90 \\
B-map  & $1.23\times10^5$  & $1.16\times10^5$  &  $0.84\times10^5$  &  $0.64\times10^5$ \\
\hline
\end{tabular}
\caption{The total SNR for different cosmological models in the ideal case. Note that, the temperature results are in line with the detection of non-Gaussianity from ISW-lensing correlation on Planck data \cite{planck-nongaussian}.}
\label{table1}
\end{table}

For comparison, we also repeat the exact same calculation for the CMB temperature anisotropy map, i.e. $T$-map, and the $E$-mode polarization map, i.e. $E$-map. Fig.\ref{fig6} shows that the non-Gaussianities in both $T$-map and $E$-map are quite small. Even though, in this figure we can also find consistent morphological characters of the cosmic lensing effects in $T$-map and $E$-map. The total SNR listed in Table \ref{table1} indicates similar results. Table \ref{table1} also shows that the non-Gaussianities of the $T$- and $E$-maps are nearly independent of $r$, since, in these two maps, the main contributions are from the scalar perturbations. In this ideal case, we find that the total signal-to-noise ratio is SNR$\sim3.5$ for the $T$-map, and SNR$\sim7.0$ for the $E$-map, being both consistent with the previous results by employing other non-Gaussianity estimators \cite{lewis}.

\begin{figure}[t]
\centerline{\includegraphics[width=16cm,height=10cm]{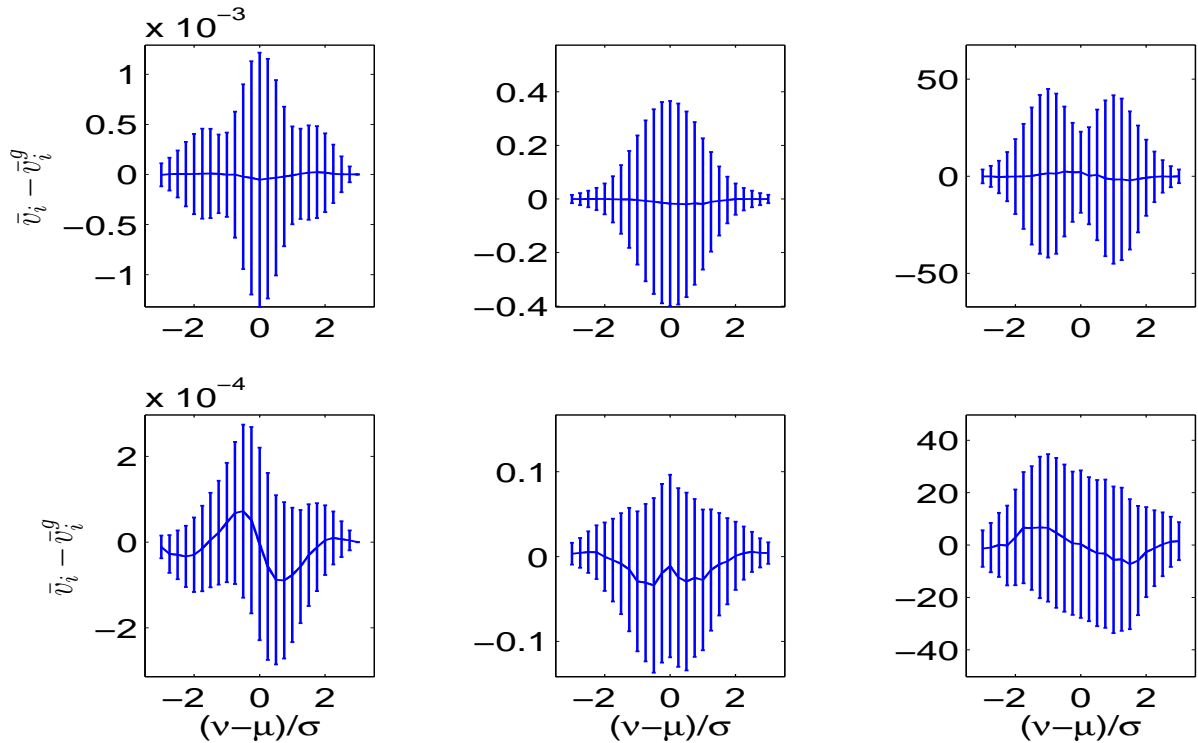}}
\caption{Upper panels denote the difference between the MFs $v_i$ in the lensed $T$-mode map and the Gaussian $T$-mode map, and lower panels denote those for $E$-map. Note that, in this figure, we have adopted the smoothing parameter $\theta_s$=$10'$ and $r=0$.}
\label{fig6}
\end{figure}



\subsection{The effect of instrumental noise}

In the real observations, the instrumental noise is inevitable and contaminate the signal of cosmological $B$-mode polarization signals in every scale. The instrumental noise becomes smaller as the experiments improve. Nowadays, although the Planck team has not yet released their observations on the $B$-mode, the instrumental noise for this observable is expected to be much larger than the primordial signal \cite{planck-white}. So, the probability for a direct detection of $B$-mode by Planck satellite is small. In the future, for both ground-based, balloon-borne and space-based experiments, the instrumental noise is expected to be reduced at least by two orders of magnitude, and become closer or even smaller than the amplitude of the lensed $B$-mode \cite{ground-based,litebird,cmbpol,core,prism,pixie}. In this paper, we consider a typical noise level of future detectors with $\Delta_P=5\mu$K-arcmin and FWHM=$10'$, where $\Delta_P$ is the constant noise per multipole and FWHM is the full width at a half maximum beam. This noise level is close to the best channel noise for the next generations of CMB experiments, both ground-based and space-based.

For the completeness of this work, we consider two cases of noise distribution in the sky. In the first case, we simply assume an homogeneous distribution of the instrumental noise. However, in practice, due to the scanning strategy and the orientations of the different horns and bolometers, the distribution of noise in the raw sky maps is correlated, anisotropic and inhomogeneous.
In the second case, we then assume an inhomogeneous noise distribution. Similar to previous works \cite{keihanen2005,ducout2012}, we treat noise effects of inhomogeneity and anisotropy by relying on hitmaps, in which each pix value represents the number of times the pixel has been observed by the detector. Our modelling of the noise in each pix of the map $i$ then reads

 \begin{equation}
 {\rm noise}(i)=\sigma_{\rm isotropic~noise}\times \mathcal{N}(0,1)\times\sqrt{\frac{\langle {\rm hitmap}\rangle}{{\rm hitmap}(i)}}.
 \end{equation}

Note that since correlations are neglected, the noise anisotropy is only partly modelled through the anisotropy of the hitmap. WMAP and Planck satellites have slightly different hitmaps in their observations. In this paper, we consider in this second case the 9-year WMAP V-band hitmap as an example to model the effect of the inhomogeneous noise in the simulations. Note that, in this step we also assume full-sky surveys.

For both noise cases considered, we superimpose the random noise maps to the simulated samples (both lensed samples and Gaussian samples), and repeat the previous calculations. The $\delta\bar{v}_i(\nu)$ ($i=0,1,2$) for the $B$-mode maps are shown in Fig. \ref{fig7} and Fig. \ref{fig8}, where we have considered the models with $r=0$ and $r=0.1$, respectively. From both figures, we clearly see that the non-Gaussianity of the $B$-mode is deeply reduced by the noise for all the MFs. {Due to the noise contaminations, the amplitudes of the deviations are reduced by a factor 5, while the error bars remain nearly the same, which are independent of the MF type or cosmological model.}

Meanwhile, for the MFs $v_0$, the main morphological character of the lensed $B$-map is kept for both noise cases. For the MF $v_1$, although the main character is preserved, the noise makes the deviation from Gaussianity in the range $\nu/\sigma\in(-1,1)$ much weaker. However, for the MF $v_2$, the main morphological character becomes different. The values of $\bar{v}_2(\nu)-\bar{v}_2^g(\nu)$ oscillate with $\nu$ value, which shows that in the ranges $\nu/\sigma\in(1,2)$ and $\nu/\sigma\in(-2,-1)$ the total integrated geodetic curvatures in the lensed maps are still smaller than those in the Gaussian maps, while in the range $\nu/\sigma\in(-1,1)$, the total integrated geodetic curvatures in the lensed maps are larger than those in the Gaussian maps. Comparing the figures for the homogeneous and inhomogeneous noise cases, we find that the average values $\bar{v}_i(\nu)$ are very close to each other in both cases, the only difference is that in the inhomogeneous noise case, the error bars of the MFs are slightly larger.

In Fig. \ref{fig9}, we show the ${\rm SNR}_{\theta_s}$ for different $\theta_s$, from which we find that the noise contaminations reduce the values of ${\rm SNR}_{\theta_s}$ by nearly or even more than one order magnitude for any $\theta_s$ case. Comparing the effects in both noise cases, we find that the deviation from Gaussian distribution in the inhomogeneous noise case is slightly smaller for any $\theta_s$. Table \ref{table2} lists the total SNR for different cosmological models, from which we find that the values of SNR decrease nearly two orders of magnitude due to the addition of the instrumental noise. Moreover, we also find that non-Gaussianity in the inhomogeneous noise case is slightly smaller than that for an homogeneous noise, which is consistent with the results shown by Figs. \ref{fig7}, \ref{fig8} and \ref{fig9}.

\begin{figure}[t]
\centerline{\includegraphics[width=18cm,height=12cm]{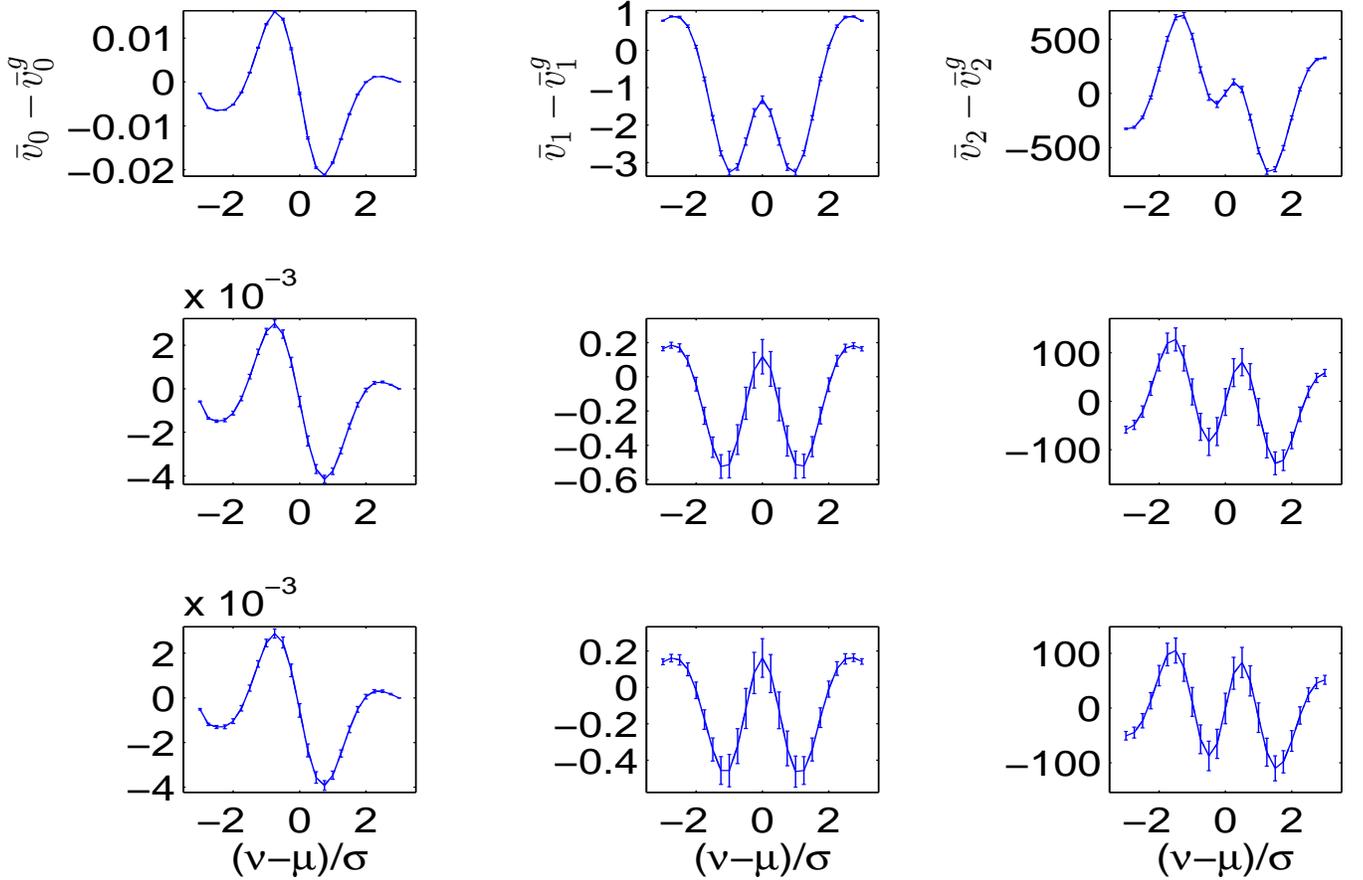}}
\caption{The difference between the MFs $v_i$ in the lensed $B$-mode map and the Gaussian $B$-mode map. The panels from upper and lower are the results for the cases without noise, with homogeneous noise and with inhomogeneous noise. In this figure, we have considered the cosmological model with $r=0$ and $\theta_s=10'$.}
\label{fig7}
\end{figure}

\begin{figure}[t]
\centerline{\includegraphics[width=18cm,height=12cm]{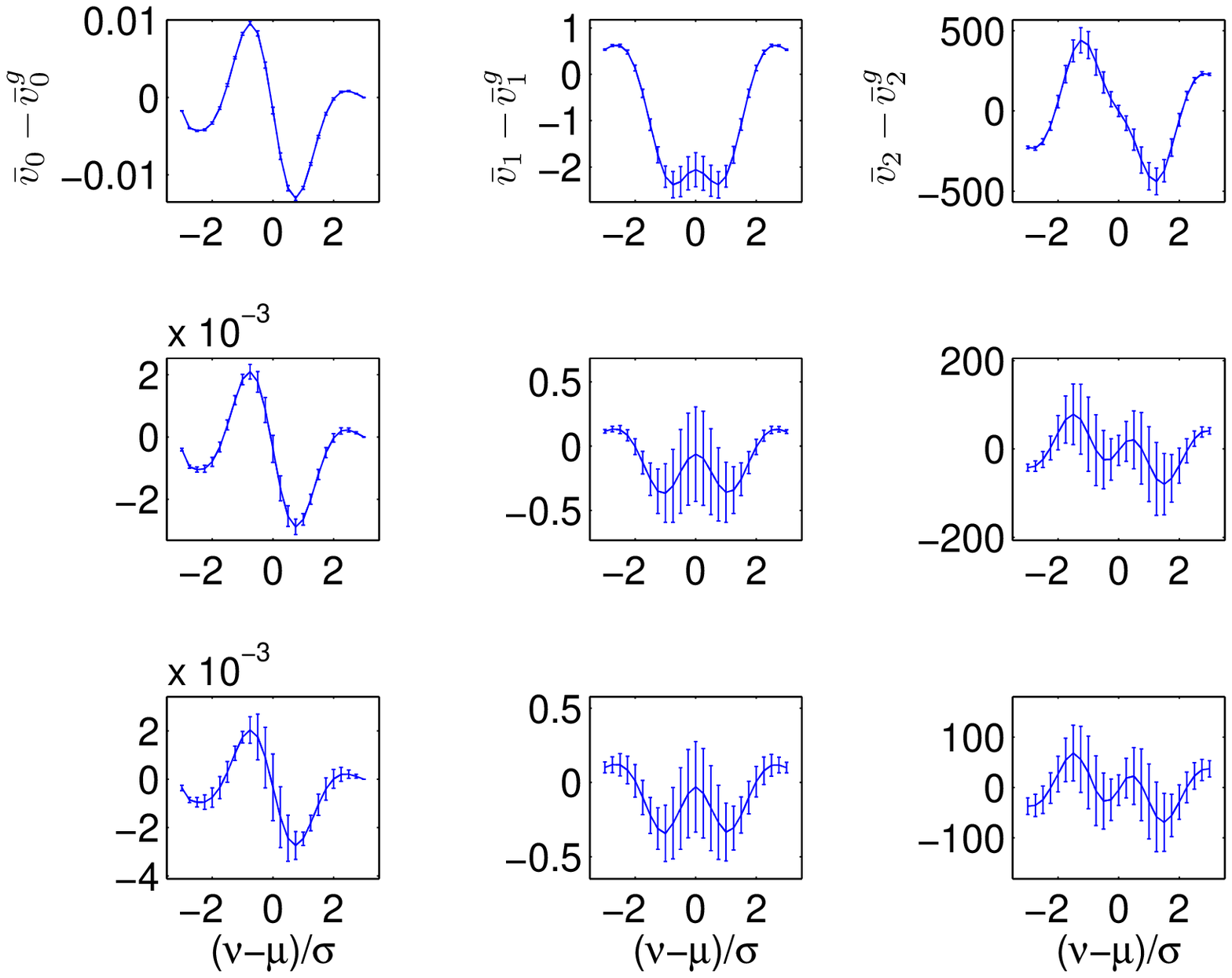}}
\caption{Same figure with Fig. \ref{fig7}, but here we consider the cosmological model with $r=0.1$.}
\label{fig8}
\end{figure}

\begin{figure}[t]
\centerline{\includegraphics[width=12cm,height=10cm]{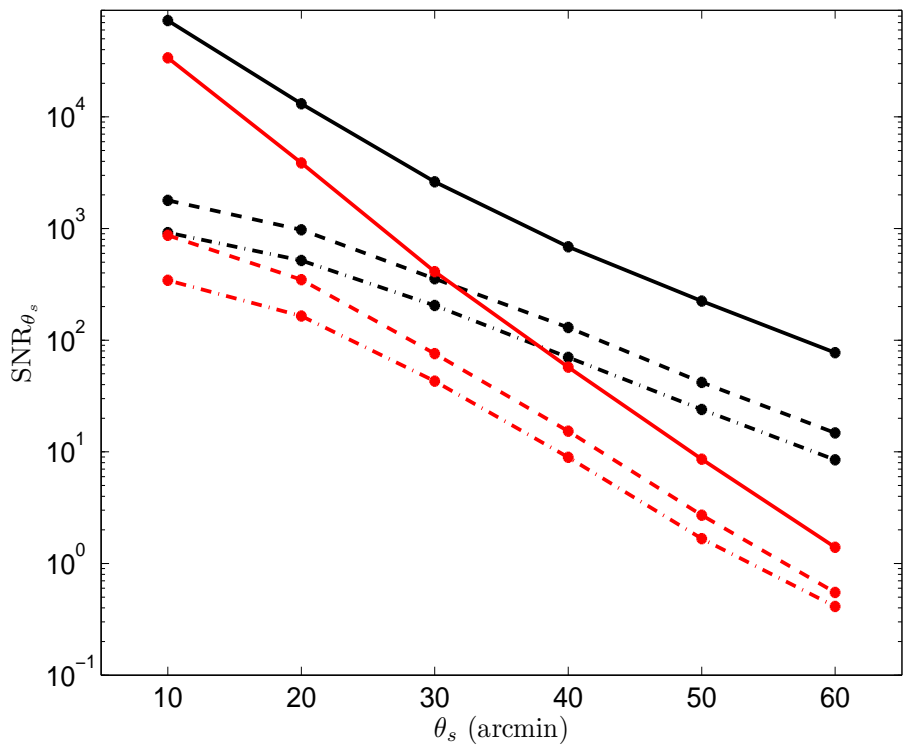}}
\caption{The SNR$_{\theta_s}$ depends on the smoothing parameter $\theta_s$ for the $B$-map. The black lines denote the results with $r=0$ and the red lines denote the results with $r=0.1$. The solid lines are for the case without noises, dashed lines are for the case with homogeneous noise, and dash-dotted lines are for that with inhomogeneous noise.}
\label{fig9}
\end{figure}

\begin{table}[!htb]
\centering
\begin{tabular}{c @{\extracolsep{2em}} c c c c}
\hline\hline
 &  $r=0$  & $r=0.01$  &  $r=0.05$  &  $r=0.1$\\
\hline
No noise & $1.23\times10^5$  & $1.16\times10^5$  &  $0.84\times10^5$  &  $0.64\times10^5$ \\
Homogeneous noise  & $3.19\times10^3$  & $2.80\times10^3$  &  $2.27\times10^3$  &  $1.57\times10^3$ \\
Inhomogeneous noise  & $1.61\times10^3$  & $1.52\times10^3$  &  $0.96\times10^3$  &  $0.63\times10^3$ \\
\hline
\end{tabular}
\caption{The total SNR for different cosmological models without noise and with noise cases.}
\label{table2}
\end{table}

\subsection{The effect of incomplete sky coverage}

The current and future CMB experiments can be divided into two different classes: One is the full-sky survey or nearly full-sky survey, which includes the space-based experiments (Planck, CMBPOL, etc), and some specifically designed ground-based or balloon-born experiments (CLASS, LSPE, etc). The other class includes most ground-based and balloon-borne experiments, in which only a small region of the sky will be surveyed, so it is impossible to get a full-sky map for the analysis. Even for the first class of experiments, we need to remove parts of the sky in order to reduce the deep contaminations caused by various foreground emissions, especially in the Galactic region. For these reasons, in practice, we always have to analyze the CMB signal in an incomplete sky.

In this paper, in order to investigate the unavoidable effect of an incomplete sky coverage in the analysis, we consider two different masks. {First of all, we use the M1 mask suggested by Planck team in 2015 (see left panel in Fig. \ref{fig10}), which is the point source and Galactic plane common mask for polarization used to produce the CMB map (version 2.00) \cite{planck-website}. \footnote{Note that this mask from the Planck collaboration is mainly intended for temperature analysis, although when foreground radiations are not included it does not change the results.}} In this mask, only the Galactic plane and some point sources of polarization are covered \cite{planck-foregrounds,planck-website}.  Similar masks may be applied to future space-based experiments data, such as the CMBPOL, COrE or LiteBird. The sky-cut factor in this case is $f_{\rm sky}=77.65\%$. In the second mask, i.e. M2 mask (right panel in Fig. \ref{fig10}), we assume the survey to be in the region defined by the Galactic coordinates: $150^{\circ}<\theta<180^{\circ}$, $-90^{\circ}<\phi<45^{\circ}$. In this region, we also exclude the pixels that have already been masked in M1 (where the polarization emissions by the point sources are strong). The Planck surveys showed that this region is one of the cleanest regions of the CMB polarization  sky (see Fig. 8 in \cite{planck-foregrounds}). This survey may mimic the observations of some ground-based experiments in South Pole. The survey area in this case is $1031$deg$^2$, which corresponds to $f_{\rm sky}=2.50\%$. {Note that, incomplete sky surveys always lead to the mixture of the $E$-mode and $B$-mode polarization, which is one of the contaminations for the $B$-mode detection \cite{ebmixture1}. Fortunately, various practical numerical methods have been developed to overcome this problem \cite{ebmixture2,ebmixture3}. Even so, the tiny leakage residuals from $E$-mode to $B$-mode are always unavoidable, which can also contaminate the intrinsic non-Gaussianity of the lensed $B$-mode polarization. In Appendix \ref{appendixA}, we discuss that if applying the method suggested by Smith and Zaldarriaga \cite{ebmixture3}, and slightly smoothing the edge of the mask (which leads to slight information loss), the leakage can be reduced to be quite small. Comparing with the case with ideal $E$-$B$ separation, we find that the influences of leakage are quite small for both power spectrum and the MFs of $B$-mode polarization. The detailed discussion on the $E$-$B$ separation and the effect on the $B$-mode statistical properties can be found in a separate future paper \cite{larissa}. But here, throughout this paper, we assume the ideal $E$-$B$ separation, i.e. ignoring the possible leakages during the separation of $E$-$B$ mixture.}


In Figs. \ref{fig11} and \ref{fig12}, we plot the $\delta\bar{v}_i(\nu)$ ($i=0,1,2$) for these two masking cases, and compare them with the full-sky case. From both figures, we find that mask M1 has little effect on any the MFs. The mean values and error bars of $v_i(\nu)$ are all very close to those in the unmasked case. In the M2 mask case, the amplitudes of the deviations $\bar{v}_i-\bar{v}_i^g$ seem slightly larger than those where M1 or no mask is being used, while the error bars become much larger due to less sky information available when using M2. In Fig. \ref{fig13}, we show the ${\rm SNR}_{\theta_s}$ for different $\theta_s$, and find the consistent results: For any $\theta_s$, the values of ${\rm SNR}_{\theta_s}$ are very close to each other in the case without using any mask and in the case considering mask M1. However, if considering mask M2, the signal-to-noise ratio becomes much smaller. The total SNRs are also calculated and listed in Table \ref{table3}, where we find that due to M2 mask, the total SNR is reduced by a factor $\sim 17$ in the model with $r=0$. For the model with $r=0.1$ the effect of M2 mask becomes relatively smaller, and the total SNR is only reduced by $\sim 11$. \emph{However, we should emphasize that even in the extreme case where the largest gravitational waves contribution is considered ($r=0.1$), inhomogeneous noise is added and M2 mask applied to the observations, the distribution of $B$-map is significantly non-gaussian. The deviation from Gaussianity is at SNR$=53$.}

\begin{figure*}[t]
\begin{center}
\includegraphics[width=5.5cm]{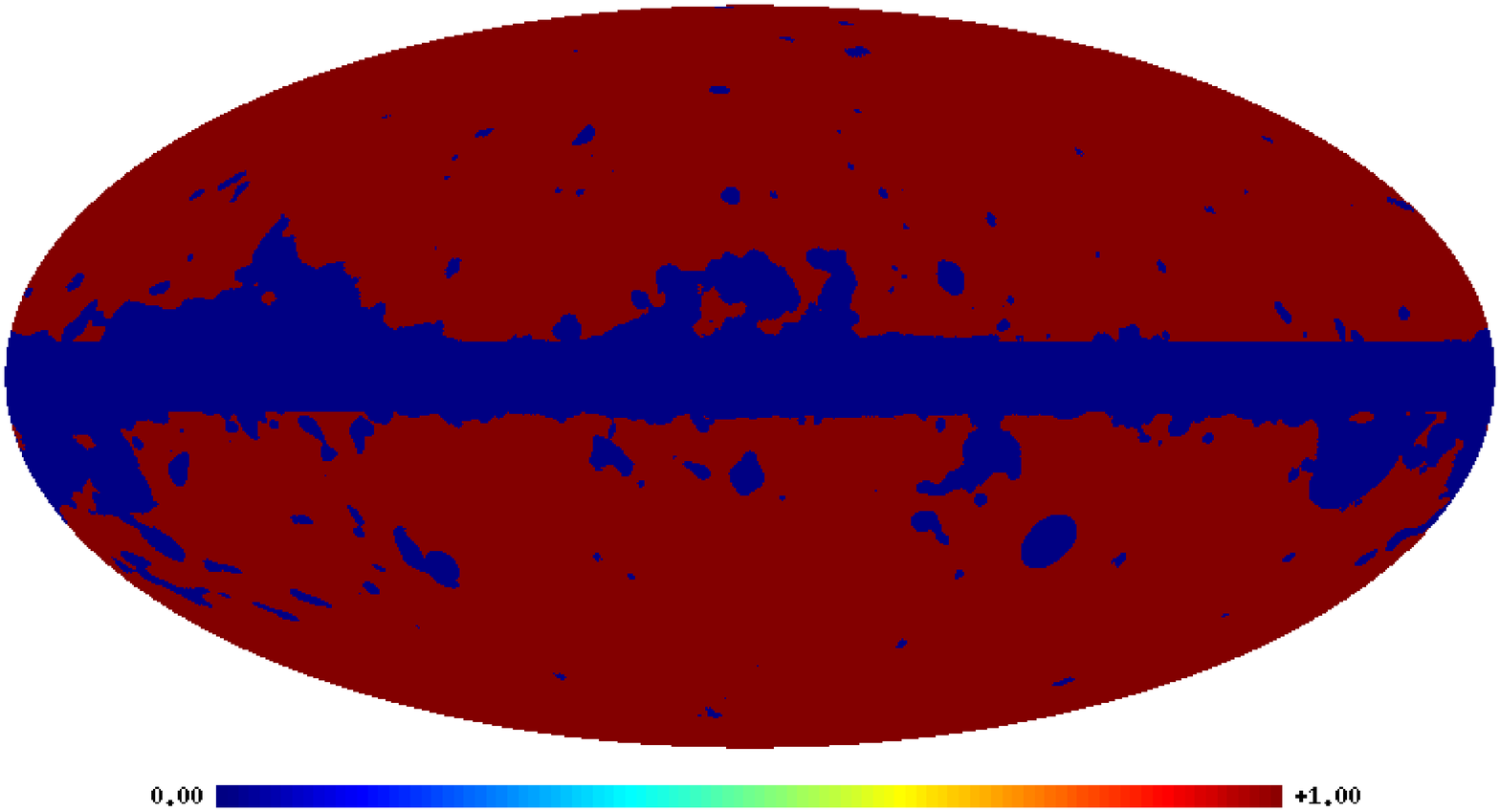}\includegraphics[width=5.5cm]{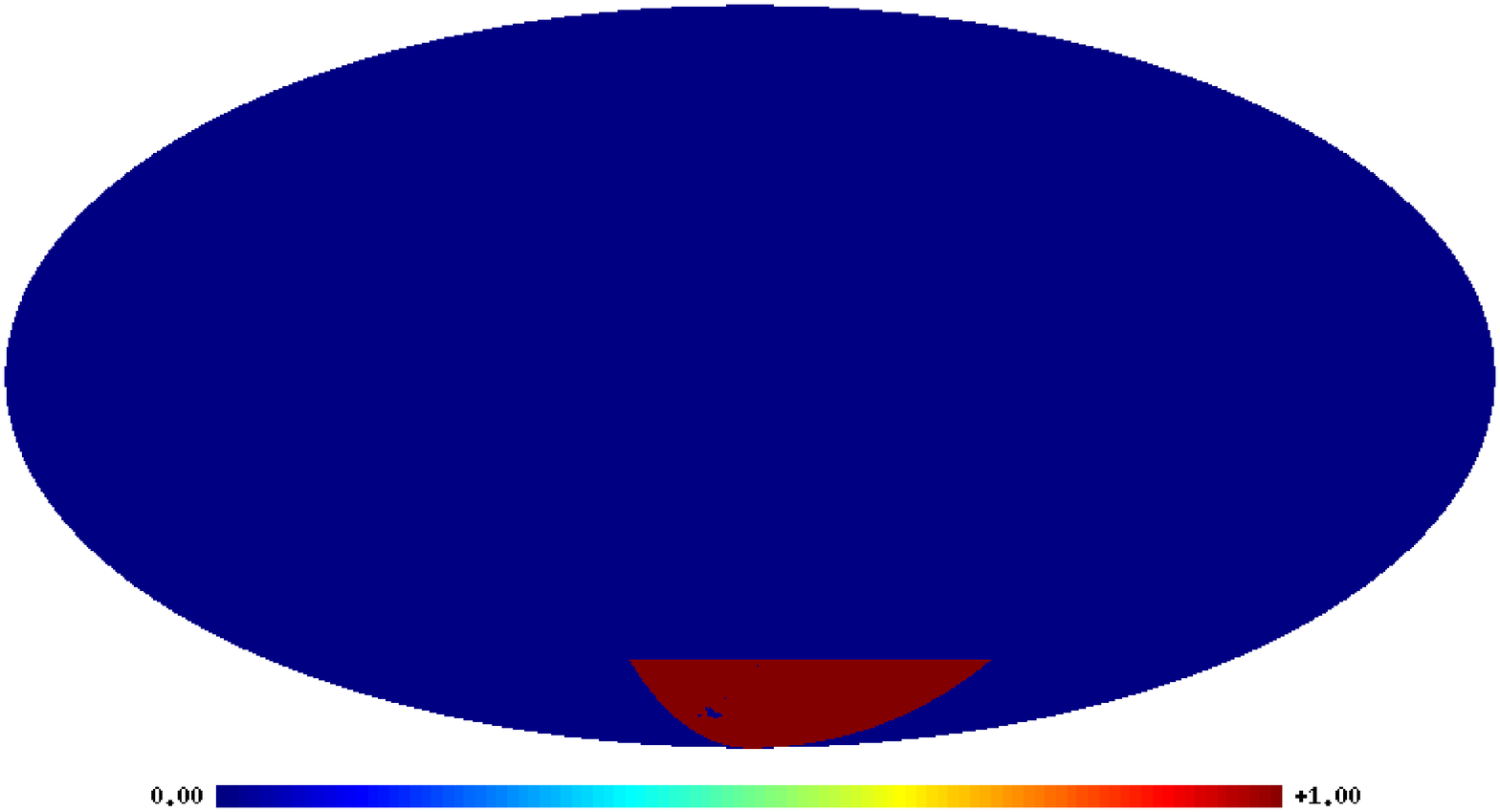}
\end{center}
\caption{The mask M1 (left) and M2 (right) used in this paper.}\label{fig10}
\end{figure*}

\begin{figure}[t]
\centerline{\includegraphics[width=18cm,height=12cm]{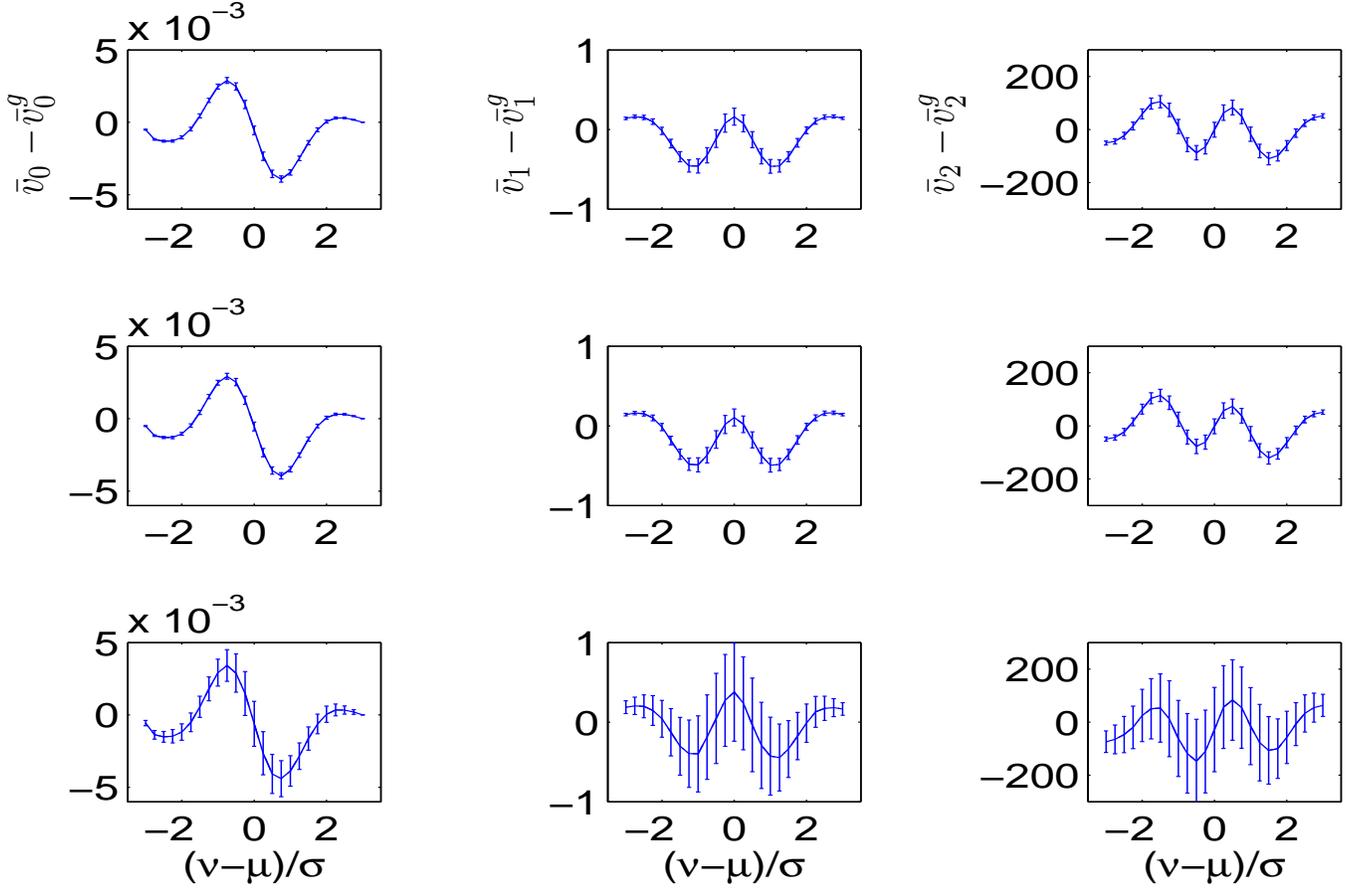}}
\caption{The difference between the MFs $v_i$ in the lensed $B$-mode map and the Gaussian $B$-mode map. The panels from upper and lower are the results for the cases with full sky, with mask M1 and with mask M2. In this figure, we have considered the cosmological model with $r=0$ and $\theta_s=10'$, and including the inhomogeneous noise.}
\label{fig11}
\end{figure}

\begin{figure}[t]
\centerline{\includegraphics[width=18cm,height=12cm]{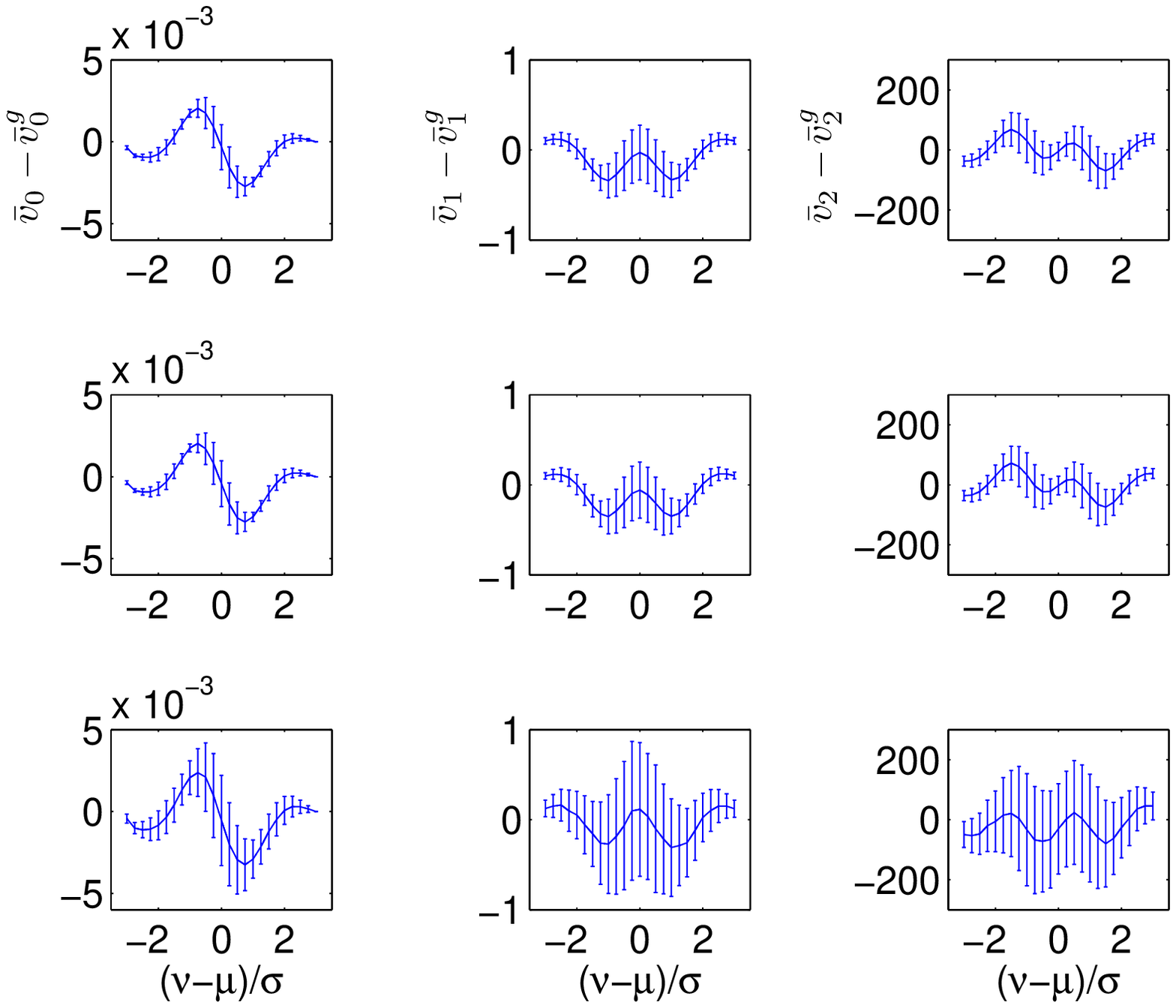}}
\caption{Same figure with Fig. \ref{fig11}, but here we consider the cosmological model with $r=0.1$.}
\label{fig12}
\end{figure}

\begin{figure}[t]
\centerline{\includegraphics[width=12cm,height=10cm]{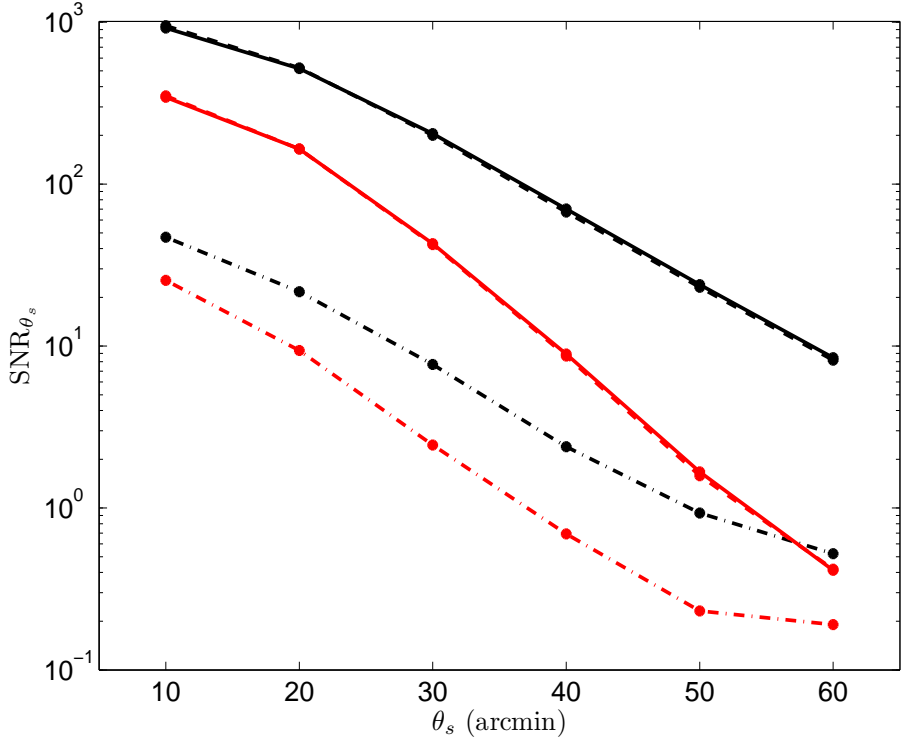}}
\caption{The values of SNR$_{\theta_s}$ depend on the smoothing parameter $\theta_s$. The black lines denote the results in the model with $r=0$ and the red lines denote the results in the model with $r=0.1$. The solid lines are the results for the case with full sky survey, dashed lines are for the case with mask M1, and dash-dotted lines are for those with mask M2. Note that, in this figure, we have included the inhomogeneous noises in our calculations.}
\label{fig13}
\end{figure}

\begin{table}[!htb]
\centering
\begin{tabular}{c @{\extracolsep{2em}} c c c c}
\hline\hline
 &  $r=0$  & $r=0.01$  &  $r=0.05$  &  $r=0.1$\\
\hline
No Mask  & $1.61\times10^3$  & $1.52\times10^3$  &  $0.96\times10^3$  &  $0.63\times10^3$ \\
Mask M1  & $1.60\times10^3$  & $1.52\times10^3$  &  $0.94\times10^3$  &  $0.64\times10^3$ \\
Mask M2  & $0.92\times10^2$  & $0.79\times10^2$  &  $0.65\times10^2$  &  $0.53\times10^2$ \\
\hline
\end{tabular}
\caption{The total SNR for different cosmological models with different masks. Note that in this table we have included the inhomogeneous noise.}
\label{table3}
\end{table}

\section{Effects of various foreground radiations}


Recent results, combining Planck satellite and BICEP2/Keck Array data, have shown that the foreground emission should be the main contamination for the future $B$-mode detection \cite{bicep2-2,planck-foregrounds}. In the high frequency channels, the main contamination is caused by dust emission, while in the low frequency channels, the dominant contamination is caused by the synchrotron radiation \cite{foregrounds}. Even in the best frequency channel, where the total foreground radiation is minimized, the amplitudes of the foregrounds are expected to be much larger than those caused by cosmological sources, including primordial gravitational waves and cosmic weak lensing \cite{bicep2-2,planck-foregrounds}.

In near future observations, by the multi-frequency channels, the foregrounds are expected to be deeply reduced by the internal linear combination or template fitting methods. However, the residuals are always unavoidable in the finial $B$-mode map. If these residuals cannot be well studied, it could mislead explanations of the data, since they might mimic the cosmological $B$-mode signal (such as what happened with BICEP2 team in 2014 \cite{bicep2-1}). It is quite difficult to distinguish them from the total observations by employing the power spectrum alone. In this section, we shall investigate if it is possible to find the imprints of the foreground residuals in the $B$-mode by employing the MF statistics, which has been partly used to the BICEP2 data in the previous work \cite{colley2014}.

{Planck recently released the CMB polarization observations $Q$ and $U$ maps at all its 7 polarized frequency bands (from 30GHz to 353GHz) \cite{planck-foregrounds,planck-website}.} Among them, the dominant component in 353GHz channel is the dust emission, while that in 30GHz channel is the synchrotron radiation. So they have been used as the templates to clean CMB polarization in the recent studies. However, it has been noticed that, the observations at these two frequency channels are also noisy as the templates of foreground radiations. In particular, in the high multipole range, the contributions of noises, including the instrumental noises, are quite large. In addition, the released polarization maps at 30GHz have the $33$ arcmin angular resolution, and those at 353GHz have the $5.0$ arcmin angular resolution \cite{planck-foregrounds}. All these make the Planck observations at these two channels bias the real foreground radiations, in particular in the high multipole range. In order to suppress the effects of these noises, we can smooth the polarization maps at both channels with the resolution parameter of $60$ arcmin. In the calculation, these smoothed maps will be used as the templates of dust emission (353GHz) and synchrotron radiation (30GHz). Here, we should emphasise that, both templates are accurate only at the large scales, i.e. in low-multipole range.

As well known, the distributions of the foreground radiations are significantly anisotropic, and both components are maximized at the Galactic plane. In order to exclude the emissions in this region, we construct a new mask, i.e. M3 mask, which covers all the regions that have already excluded in M1 mask. In addition, as a conservative estimation, we also mask the pixels where the origin values of $Q$ or $U$ at 353GHz are larger than $10\mu$K$_{\rm RJ}$, and the pixels where the origin values of $Q$ or $U$ at 30GHz are larger than $20\mu$K$_{\rm RJ}$. This mask is presented in Fig. \ref{fig14} (left panel), and it has the cut-sky factor $f_{\rm sky}=69.8\%$.

Based on these knowledge, we roughly evaluate the residual foreground maps by the following steps:

(1) For the dust component, we construct the $E$- and $B$-mode maps at 353GHz frequency channel from the observed $Q$ and $U$ map by the standard procedures. Similarly, the $E$ and $B$-mode maps of synchrotron radiation component at 30GHz channels are also constructed. Then, we smooth these $B$-mode maps with the resolution parameter of $60$ arcmin.

(2) It is known that the dependence of dust $B$-mode power spectrum $D_{\nu}$ on the frequency $\nu$ satisfies the following simple parametrization, $D_{\nu}\propto(W_{\nu}^{d})^2$, where $W_{\nu}^{d}=(\frac{\nu}{\nu_d})^{1+\beta_d}\frac{1+e^{h\nu_d/kT}-1}{1+e^{h\nu/kT}-1}$. The parameters are $\beta_d=1.59$, $\nu_d=353$GHz and $T=19.6$K \cite{planck-foregrounds,planck-foreground2,foregrounds}. For the synchrotron radiation spectrum, the parametrization $S_{\nu}$ is $S_{\nu}\propto(W_{\nu}^{s})^2$, where $W_{\nu}^{s}=(\frac{\nu}{\nu_s})^{\beta_s}$. The parameters are $\beta_s=-2.9$, $\nu_s=30$GHz \cite{planck-foregrounds,planck-foreground2,foregrounds}. In our calculations, we assume that the useful channel for cosmological studies is at 150GHz. Thus, the dust (synchrotron radiation) $B$-mode map can be evaluated for 150GHz from the 353GHz map (and the 30GHz map) by employing the above parameterizations.

(3) We mask the smoothed $B$-maps at 150GHz channel by applying the M3 mask, which are presented in the middle and right panels of Fig. \ref{fig14}, respectively. Note that we have converted the antenna temperature $T_A$ to the thermodynamic temperature in the calculations. {The root mean square (RMS) of the middle panel (dust emission) in the unmasked region is $6.67\mu$K. While for the synchrotron radiation at 150GHz (right panel), it is much smaller, i.e. RMS of the right panel in the unmasked region is $0.0368\mu$K.}

\begin{figure}[t]
\begin{center}
\includegraphics[width=5.5cm]{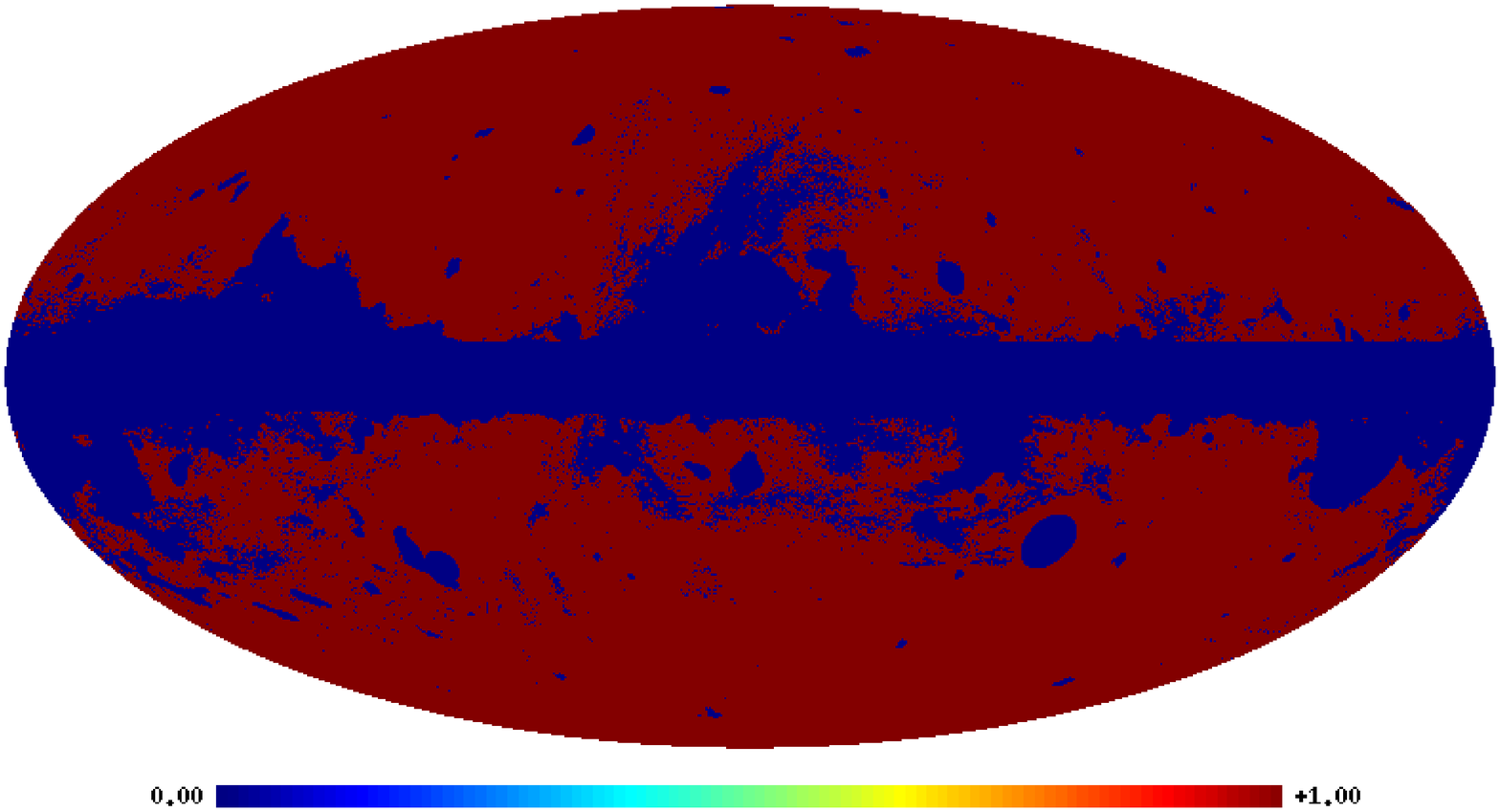}
\includegraphics[width=5.5cm]{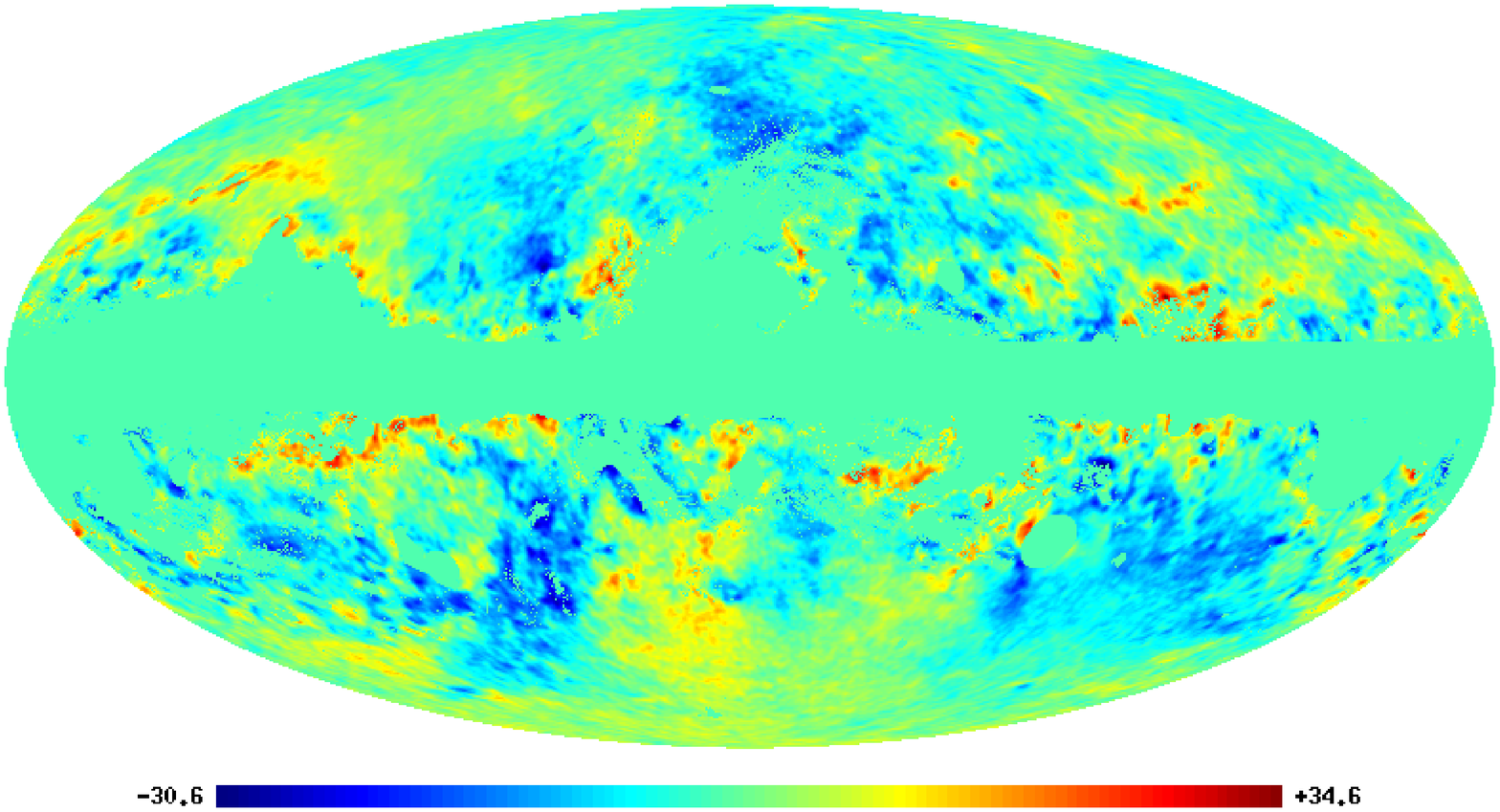}
\includegraphics[width=5.5cm]{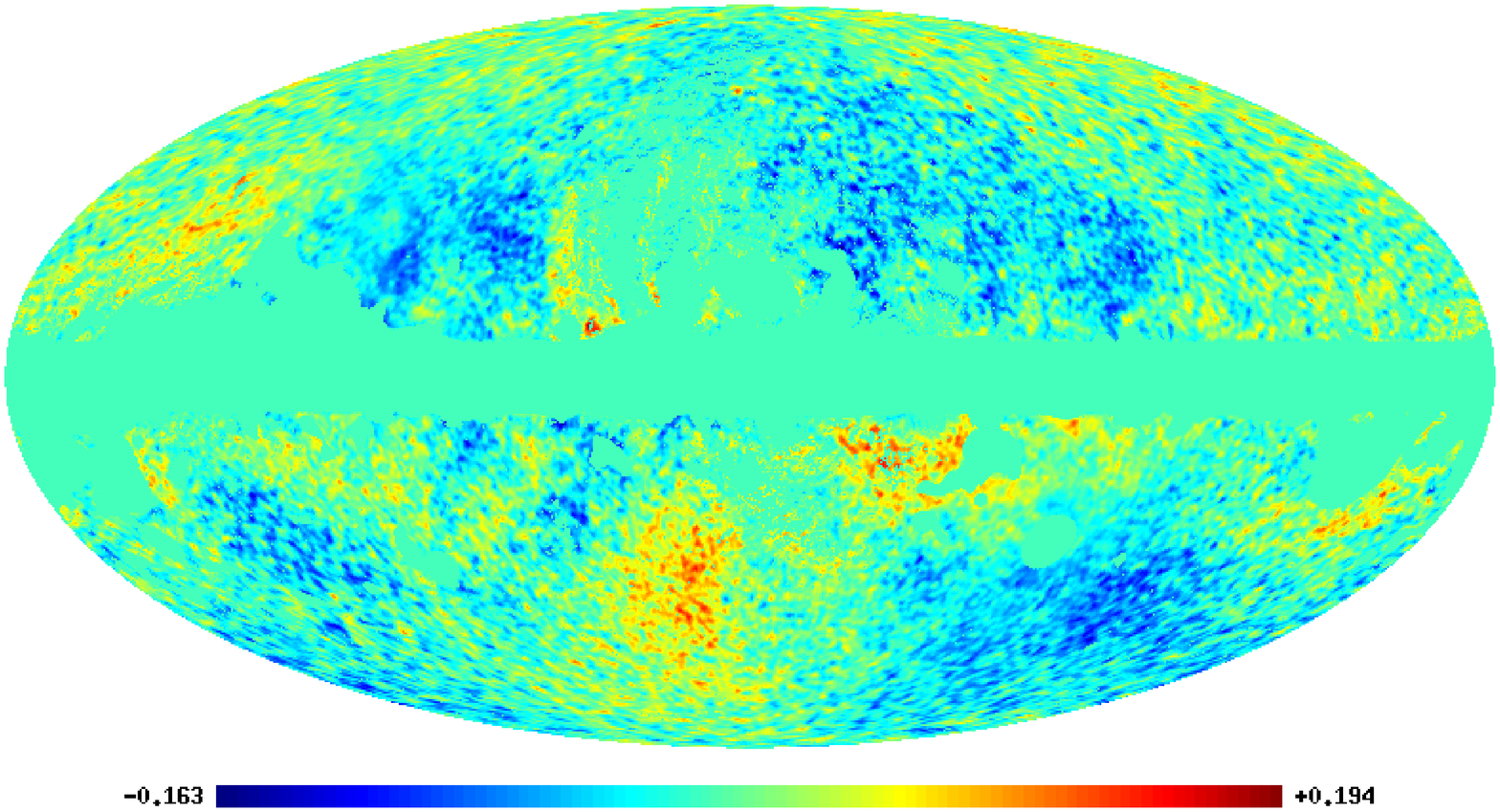}
\end{center}
\caption{The M3 mask (left) used in this paper. The middle and right panels are the smoothed $B$-mode polarization maps (at $150$GHz frequency channel) caused by dust emission and synchrotron radiation respectively, which have been masked by M3 mask. Note that, the units in the middle and right panels are $\mu$K.}\label{fig14}
\end{figure}

(4) In order to estimate the residual foregrounds left in the CMB map after templates fitting and components separation, we define the residual factor $\sigma_{\rm dust}$ and $\sigma_{\rm syn}$ for dust emission and synchrotron radiation, respectively. Actually, it is impossible to know exactly the morphology and distribution of the residuals. However, we expect that the morphologies of the residual maps should be similar to those of the foreground radiation maps. So, in our calculation, we evaluate the residual maps (middle and right panels in Fig. \ref{fig14}) by simply multiplying the full foreground maps by the residual factor $\sigma$.

By applying the pseudo-$C_{\ell}$ method (see Appendix \ref{appendixC}) \cite{method4} to the masked $B$-maps in Fig. \ref{fig14}, we construct the estimators for the $B$-mode power spectra. {As we have emphasized above, due to the smoothing of the maps, these power spectra are the biased ones for the real dust and synchrotron emissions, which are accurate only at the low multipole range $\ell\lesssim100$.} As anticipated, we find that the amplitude of the dust contamination is much larger than that of cosmological signals. While for the synchrotron radiation component, the amplitude of $B$-mode is much smaller than for the dust component, and it seems even smaller than the cosmological components if the gravitational waves contribution is large enough. So, as expected, in this frequency channel, the dominant foreground contamination will be the dust emission. The power spectra of foreground residuals can be obtained by simply multiplying by factor $\sigma^2$. From Fig. \ref{fig15}, we find that if applying the M3 mask, and considering the cosmological model including weak lensing and gravitational waves with $r=0.1$, the residuals of the synchrotron radiation at 150GHz is always subdominant, which is independent of the $\sigma_{\rm syn}$ value. While, for the dust emission, the residuals will become subdominant if the $\sigma_{\rm dust}\lesssim3\%$.

\begin{figure}[t]
\centerline{\includegraphics[width=13cm,height=10cm]{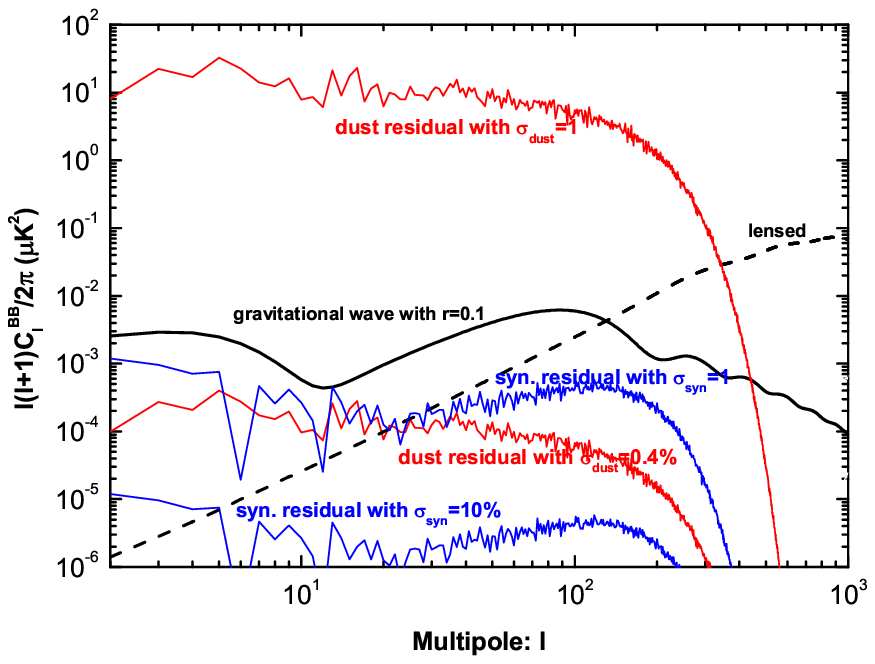}}
\caption{ The $BB$ power spectra for the residuals of dust emission (red lines) and synchrotron radiation (blue lines) at 150GHz. For comparison, in this figure we also plot the lensed $BB$ spectrum (black dashed line) and the $BB$ spectrum caused by primordial gravitational waves with $r=0.1$ (black solid line). In order to reduce the effect of noises, for both foreground emissions, we have applied the smoothing with the resolution parameter of $60$ arcmin. (see main text for the details).}
\label{fig15}
\end{figure}

~

To study the effects of these residuals, we superimpose the constructed foreground maps to each simulated lensed $B$-mode map. Consequently, we can derive the MFs for the contaminated maps by repeating the computations mentioned in Sec. IV. Fig. \ref{fig16} shows the results for the case where $\sigma_{\rm dust}=1\%$ and $\sigma_{\rm syn}=0\%$. We clearly see that all three MFs are definitely different from those when no foregrounds are considered. This result is expected, since, from Fig. \ref{fig15}, we can notice that the amplitudes of the contaminated $B$-mode and the cosmological $B$-mode are comparable for the case with $\sigma_{\rm dust}=1\%$. We quantify the effects of the foreground residuals by the following $\chi^2$ quantity,
 \begin{equation}
 \chi^2 = \sum_{aa'}\left[\bar{v}_{a}-\bar{v}^{\rm no}_{a}\right]\Sigma_{aa'}^{-1}
 \left[\bar{v}_{a'}-\bar{v}^{\rm no}_{a'}\right],
 \end{equation}
where $v_{a}$ is the MF calculated from the contaminated map, and $v_{a}^{\rm no}$ is the corresponding MF $v_a$ in the foreground free map. $\Sigma$ is the covariance matrix for $v_a$.
Here $a$ and $a'$ denote the binning number of threshold value $\nu$, the different kinds of MF $i$ and the smoothing scale parameter $\theta_s$.

\begin{figure}[t]
\centerline{\includegraphics[width=18cm,height=18cm]{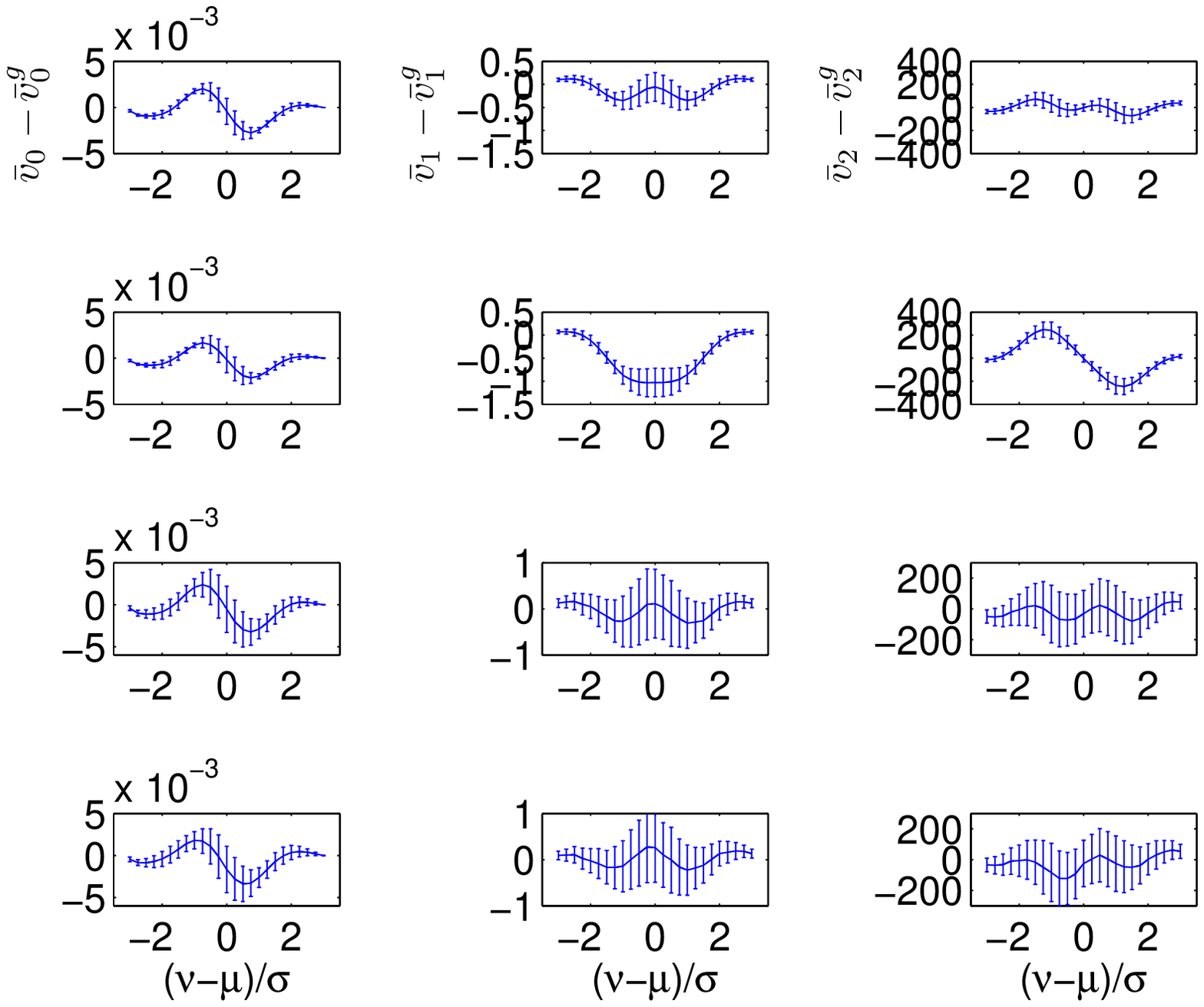}}
\caption{The difference between the MFs $v_i$ in the $B$-mode map and the Gaussian $B$-mode map.The upper two rows show the results when employing the M3 mask. Among them, in the first row we consider the case without foreground residuals, and in the second row we consider the case including $1\%$ dust residual at 150GHz. The lower two rows show the results when employing the M2 mask. Among them, in the third row we consider the case without foreground residuals, and in the fourth row we consider the case including $1\%$ dust residual at 150GHz. In this figure, we have considered the cosmological model with $r=0.1$, applying the smoothing parameter $\theta_s=10'$, and including the inhomogeneous noise.}
\label{fig16}
\end{figure}

In Table \ref{table4}, we list the $\chi^2$ values when M3 mask is applied. In this calculation, we have included inhomogeneous noise to the simulations. We find that for any given $\sigma_{\rm dust}$ or $\sigma_{\rm syn}$, the values of $\chi^2$ decrease when the gravitational waves contribution increases, which is understandable: Increasing the gravitational waves contribution, also increases amplitude of the total cosmological $B$-mode, especially in the large angular scales. In such case, the relative proportion of the foreground residuals decreases, which makes the difference between polluted map and clean map smaller. From this Table, we find that when only dust residuals are taken into account, and for $\sigma_{\rm dust}>0.3\%$ ($\sigma_{\rm dust}>0.4\%$), these differences can be detected by MF analysis at more than $\sim2\sigma$ ($\sim5\sigma$) confidence level. {Note that in these two cases, the power spectrum of foreground residuals is nearly two orders of magnitude smaller than those of the signals (see Fig. \ref{fig15})}. However, the effect of the synchrotron radiation in this channel is quite small. Even if $\sigma_{\rm syn}=10\%$, the difference between the polluted map and clean map can be detected by the MF analysis at only $\lesssim 2\sigma$ level.

In order to mimic the conditions of the future small-sky surveys ({e.g.} the future ground-based experiments), we consider another case, in which the M3 mask is replaced by the M2 mask. {\it Note that, as we have mentioned, the unmasked region is one of the cleanest regions for the $B$-mode detection \cite{planck-foregrounds}.} We repeat the computations and list the corresponding $\chi^2$ values in Table \ref{table5}. We find that in this case, the effects of the foreground residuals are not so important. There are two reasons for this: First, the unmasked region in M2 mask is one of the cleanest region for the CMB polarization \cite{planck-foregrounds}. Second, there is less sky information available when this mask is used, which makes the foregrounds detection more difficult. Nevertheless, in this case, if the dust residual is beyond $1\%$, the difference between polluted map and clean map can be detected by the MF analysis at more than $\sim4\sigma$ confidence level.

\begin{table}[!htb]
\centering
\begin{tabular}{c @{\extracolsep{2em}} c c c c}
\hline\hline
 &  $r=0$  & $r=0.01$  &  $r=0.05$  &  $r=0.1$\\
\hline
$\sigma_{\rm dust}=0.1\%$  & $7.36$  & $1.92$  &  $1.71$  &  $1.63$ \\
$\sigma_{\rm dust}=0.2\%$  & $84.65$  & $6.42$  &  $2.05$  &  $1.74$ \\
$\sigma_{\rm dust}=0.3\%$  & $360.41$  & $22.97$  &  $4.13$  &  $2.01$ \\
$\sigma_{\rm dust}=0.4\%$  & $1003.16$  & $72.09$  &  $9.07$  &  $4.92$ \\
$\sigma_{\rm dust}=0.5\%$  & $2127.13$  & $167.32$  &  $20.34$  &  $8.99$ \\
$\sigma_{\rm dust}=1.0\%$  & $15932.43$  & $2725.01$  &  $332.27$  &  $97.09$ \\
$\sigma_{\rm syn }=10.0\%$  & $2.17$  & $1.95$  &  $1.64$  &  $1.53$ \\
\hline
\end{tabular}
\caption{$\chi^2$ values for the case with inhomogeneous noise and applying the M3 mask. The last row shows the results with only synchrotron radiation, and the other rows show those with only dust contamination.}
\label{table4}
\end{table}

\begin{table}[!htb]
\centering
\begin{tabular}{c @{\extracolsep{2em}} c c c c}
\hline\hline
 &  $r=0$  & $r=0.01$  &  $r=0.05$  &  $r=0.1$\\
\hline
$\sigma_{\rm dust}=0.5\%$  & $3.41$  & $2.29$  &  $1.90$  &  $1.89$ \\
$\sigma_{\rm dust}=1.0\%$  & $15.02$  & $8.31$  &  $4.02$  &  $3.87$ \\
$\sigma_{\rm dust}=5.0\%$  & $1027.51$  & $686.79$  &  $219.76$  &  $225.37$ \\
~$\sigma_{\rm dust}=10.0\%$  & $2479.43$  & $1702.65$  &  $699.31$  &  $550.17$ \\
$\sigma_{\rm syn }=10.0\%$  & $1.71$  & $1.50$  &  $1.51$  &  $1.42$ \\
\hline
\end{tabular}
\caption{Same with Table \ref{table4}, but here we have considered the M2 mask, instead of the M3 mask.}
\label{table5}
\end{table}

The $B$-mode polarization generated by foreground residuals is different from that caused by the cosmological sources at several aspects. Since the residuals are directly related to the original foreground radiation, we expect them to have inhomogeneous, anisotropic, non-random distributions. Furthermore, the power spectra of the residual $B$-modes should be also different from the cosmological ones. In the previous discussion, we find that the deviations of the MFs caused by the foreground residuals, in comparison with the the MFs for the clean map, are quite promising for the detection. However, it is still possible that this deviation is mainly caused by the extra power spectrum contribution from the foregrounds, and the effects of the other morphological properties are very small. If this is the true case, the effects of foregrounds and cosmological parameters (in general, changing the cosmological parameters can also induce the changing the CMB power spectrum) become degenerate, and the detection of residual imprints of foreground by the MFs is actually impossible.

In order to investigate this issue, we do the comparison for the following two cases: In Case I, we randomly simulate 1000 $B$-mode maps by considering the cosmological model with $r=0.1$ (including the lensing effect). Then, we superimpose the residuals of the dust emission with $\sigma_{\rm dust}=0.3\%$, and employ the M3 to mask the combined $B$-maps. Then, we calculate the mean values $\bar{v}_i(\nu)$ and covariance matrix $\Sigma$ of MFs in these maps. In Case II, we do the exact same calculation as in Case I, but the residual of the dust emission component is replaced by a simulated random Gaussian field, which has the exact same power spectrum with the dust residuals. The mean values of MFs in Case II are denoted as $\bar{v}^c_{i}(\nu)$. We quantify the difference between these two cases by the $\chi'^2$ defined as follows,
 \begin{equation}
 \chi'^2 = \sum_{aa'}\left[\bar{v}_{a}-\bar{v}^c_{a}\right]\Sigma_{aa'}^{-1}
 \left[\bar{v}_{a'}-\bar{v}^c_{a'}\right],
 \end{equation}
where $a$ and $a'$ denote the binning number of threshold value $\nu$, the different kinds of MF $i$ and the smoothing scale parameter $\theta_s$. We obtain that $\chi'^2=3.29$. If $\sigma_{\rm dust}=0.4\%$ is used, we have $\chi'^2=6.01$. We notice that although the power spectra in these two cases are the same, their morphological characters are definitely different, which have important effects on the MFs. This result is easy to understand: For example, due to the anisotropic distribution of the foreground radiation in the sky, the residuals have larger power spectrum in regions close to the mask edges, and smaller power spectrum in regions far away from the mask edges. So, when we calculate the MFs with different smoothing scale parameters $\theta_s$, the amplitudes of foreground residuals are different. A larger $\theta_s$ means a deeper smoothing, which corresponds to a cleaner map. Thus, the morphological distribution of the foregrounds is encoded by our choices of several $\theta_s$ (from $10'$ to $60'$) values in the calculations.

\section{Conclusions}

The CMB radiation is completely described by the scalar fields $T$ and $E$, and the pseudo-scalar field $B$. In the standard cosmological model, according to the linear theory, $T$ and $E$ are excellent random Gaussian fields, and the non-Gaussianities are expected to be very weak. So, the two-point correlation functions, or equivalently the angular power spectrum in harmonic space, are enough to describe the total statistical properties of the fields. However, the $B$-mode should be a strongly non-gaussian field, due to the important contribution of cosmic weak lensing effect. So, in addition to the two-point correlation function, the other statistics are also very important, because they contain the complementary information of the $B$-mode field, and should be helpful to constrain cosmological parameters or separate different components of the total map.

In this paper, we employed the MFs to study the morphological characters of the lensed $B$-mode polarization map. By comparing it with purely Gaussian fields, we find that the $B$-map definitely deviates from Gaussianity. For both nearly full-sky surveys and partial sky surveys, the non-Gaussianity can be detected at very high confidence level. From the MF analysis, we also find that, comparing with the Gaussian fields, the lensed maps seem to have more pixels in the regions $|\nu|\in(1\sigma,2\sigma)$. Moreover, hot and cold spots have the tendency to form smooth clusters, which is consistent with general effect of the gravity on matter distributions.

As an application of the MF analysis, we have investigated the imprints of the foreground residuals in the $B$-mode map. We find that for the potential satellite CMB experiments, if more than $0.4\%$ of dust emission is left in the map as residuals, its effect can be detected by MF analysis. Even for the potential ground-based experiments with partial sky surveys, the dust emission can be detected if more than $1\%$ of residuals are left in $B$-map. In addition, we find that the influence of the foreground residuals is not simply caused by their extra power spectra, and the special morphology of the foreground radiations may play the crucial role.

~

~

\section*{Acknowledgements}
We acknowledge the use of the Planck Legacy Archive (PLA). Our data analysis made the use of HEALPix
\cite{healpix}, CAMB \cite{camb} and LensPix \cite{lenspix}.
This work is supported by Project 973 under Grant No. 2012CB821804, by NSFC No. J1310021, 11173021, 11322324, 11421303 and project of Knowledge Innovation Program of Chinese Academy of Science.

\appendix

\section{The skewness and kurtosis statistics for the lensed $B$-mode polarization \label{appendixB}}

In this Appendix, we apply the one-point statistics: skewness ($S$) and kurtosis ($K$), to analyse the statistical properties of the CMB $B$-mode polarization maps. For simplification, we do not include the contribution from any kind of noise or use any mask in this analysis. In probability theory, skewness is a measure of the asymmetry of the probability distribution of a real-valued random variable about its mean. From the MF analysis above, especially the MF $v_0$, we know that in the lensed $B$-mode map, the distribution of the points are symmetric around the mean value $\mu$ (see for instance the left panels of Fig. \ref{fig2}). So, we expect that the values of the skewness for the lensed $B$-mode maps are close to zero. Similar to the MF analysis above, we simulate 1000 independent lensed $B$-mode sample using the LensPix package. For each sample, we calculate the skewness, and present the histogram of these $S$'s in Fig. \ref{figb1} (upper left panel). As expected, we find that their values are all very close to zero. The mean value and the standard deviation of these 1000 $S$'s are $0.0000\pm0.0047$. For comparison, we simulated 1000 Gaussian samples, which have the exact same power spectrum of the lensed $B$-mode sample. We then calculate the skewness of each sample. The histograms of these statistics are presented in Fig. \ref{figb1} (upper right panel). From this figure, we find that the distribution of skewness in the lensed $B$-map is similar to that in the Gaussian maps. The only difference is that in the Gaussian maps, the width of the distribution is slightly narrower. If increasing the value of the smoothing scale parameter $\theta_s$ and/or the contribution of gravitational waves, from Figs. \ref{figb2}, \ref{figb3} and Table \ref{tableb1}, we find that the distribution of the statistics become broader, while their mean values are all close to zero. This analysis shows that skewness cannot well describe the non-Gaussianity of the lensed $B$-mode polarization.

The situation is different for the kurtosis statistic. From histograms of $K$'s derived from the 1000 lensed $B$-mode maps, we find that the values of the kurtosis are all around of 1 (see lower left panels in Figs. \ref{figb1}-\ref{figb3}), against 0 for the Gaussian case (see lower right panels in Figs. \ref{figb1}-\ref{figb3}). This shows that the expectation value of the kurtosis in the lensed $B$-maps is much larger than that in the Gaussian maps. Since kurtosis is a measure of whether the data is peaked or flat relative to a normal distribution, the larger kurtosis indicates that the one-point distribution of lensed $B$-map is relatively flat than that of the Gaussian distribution, which is also consistent with the MF analysis, in particular, the $v_0$ functional. With the increasing of $\theta_s$ and/or $r$, the mean values of the kurtosis decrease (see Figs. \ref{figb1}-\ref{figb3} and Table \ref{tableb2}), which indicates that the non-Gaussianity becomes smaller. In order to quantify the non-Gaussianity, we define the quantity ${\rm SNR}_{\theta_s}=\Delta_K/\sigma_K$, where $\Delta_K$ is the difference between the mean value of kurtosis in the lensed $B$-maps and the corresponding one in the Gaussian $B$-maps, and $\sigma_K$ is the standard deviation of kurtosis in the lensed $B$-maps. Using the results listed in Table \ref{tableb2}, we get that ${\rm SNR}_{\theta_s}=63.5$ for the case with $\theta_s=10'$ and $r=0$. If $\theta_s=60'$ and $r=0$, we have ${\rm SNR}_{\theta_s}=3.9$. While for the case with $\theta_s=10'$ and $r=0.1$, we derive ${\rm SNR}_{\theta_s}=27.1$. These show that kurtosis is a good statistic to describe the non-Gaussianity of the lensed $B$-maps. However, comparing with the results derived from the MF analysis, we also find that the sensitivity of kurtosis is worse than the MFs. Here, we would like to emphasize that the kurtosis and MFs encode different non-gaussian information, so it its better to use both statistics in the future analysis.

\begin{figure}[t]
\centerline{\includegraphics[width=12cm]{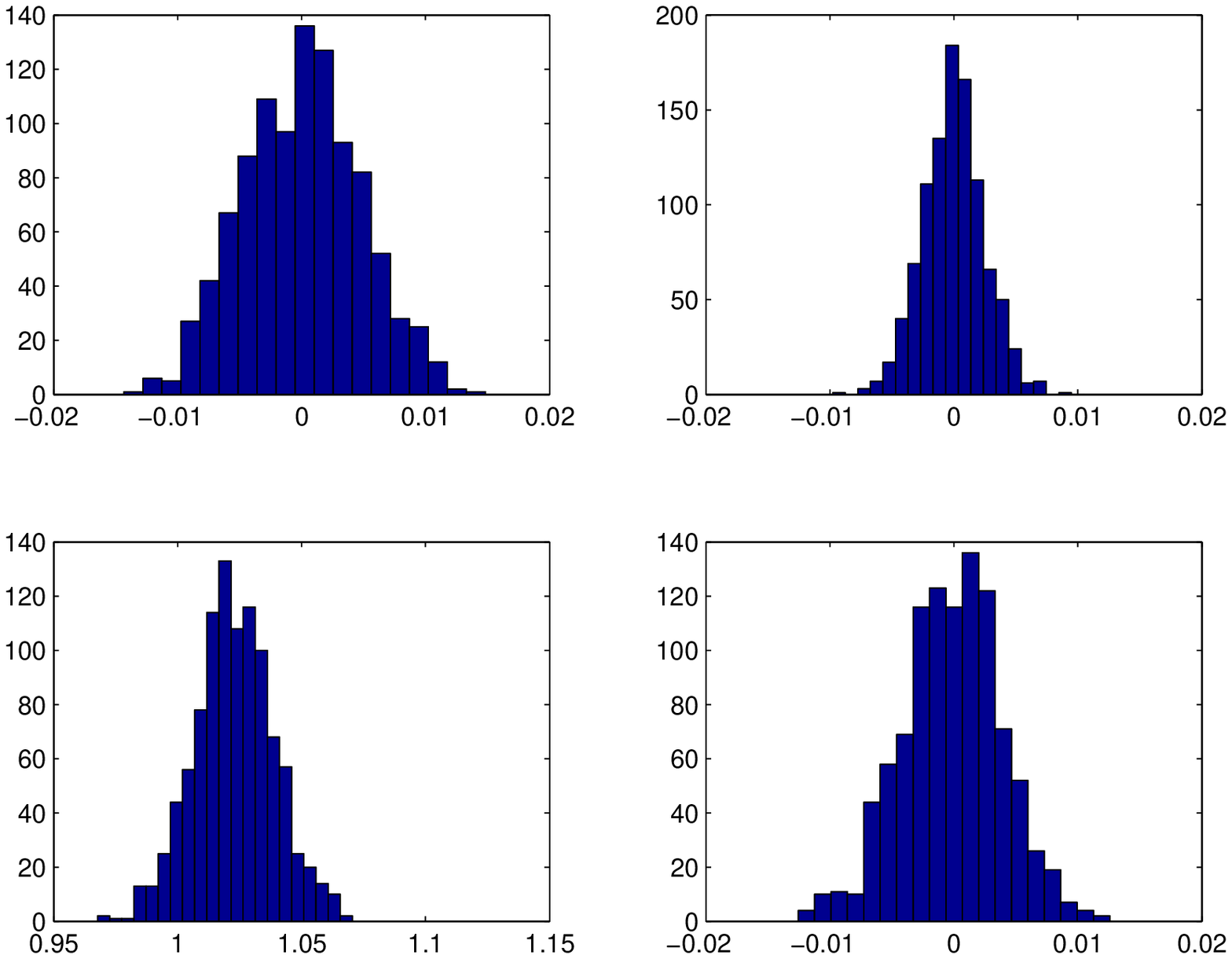}}
\caption{The upper panels show the histogram of the skewness statistic. The left one is derived from 1000 lensed $B$-mode realizations, and the right one is derived from 1000 Gaussian realizations. The lower panels show the histogram of the kurtosis statistic. The left one is derived from 1000 lensed $B$-mode realizations, and the right one is derived from 1000 Gaussian realizations. In this figure, we have considered the model without gravitational wave, i.e. $r=0$, and adopted the smoothing scale parameter $\theta_s=10'$.}
\label{figb1}
\end{figure}

\begin{figure}[t]
\centerline{\includegraphics[width=12cm]{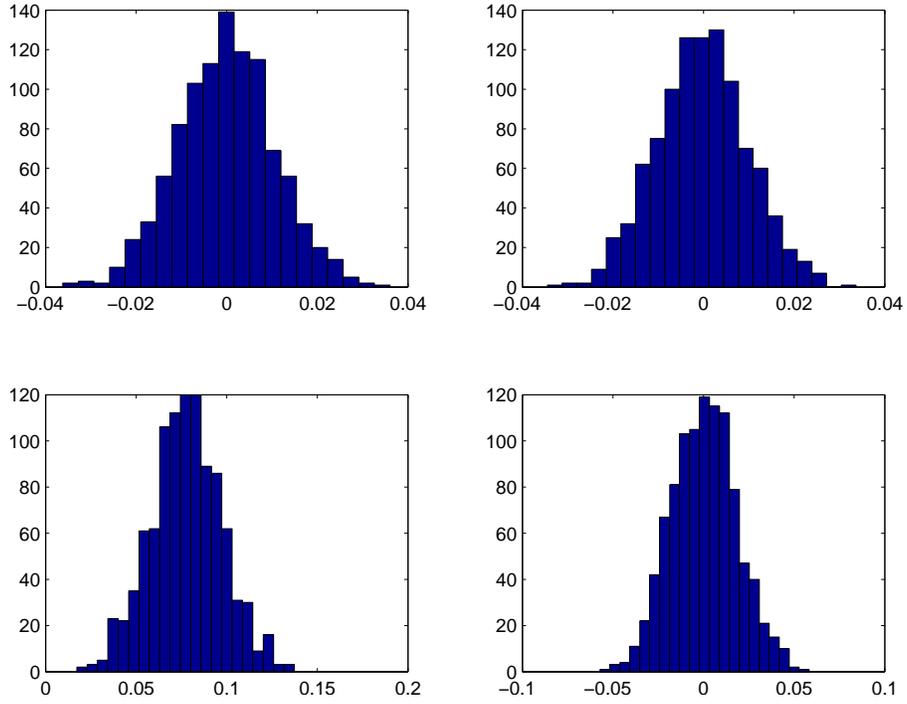}}
\caption{Same figure with Fig. \ref{figb1}, but here we consider the smoothing scale parameter $\theta_s=60'$.}
\label{figb2}
\end{figure}

\begin{figure}[t]
\centerline{\includegraphics[width=12cm]{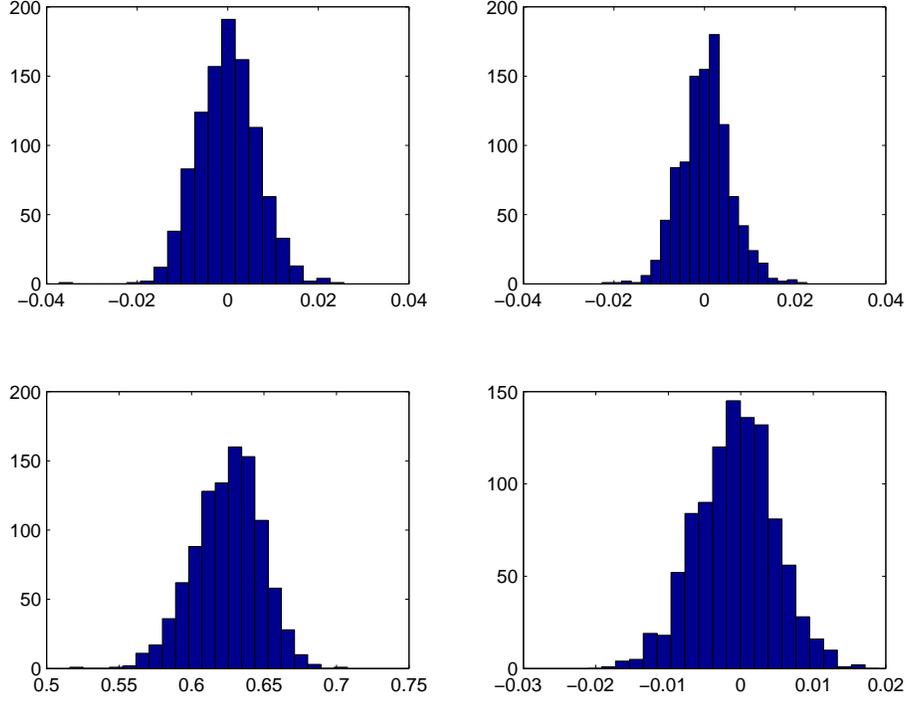}}
\caption{Same figure with Fig. \ref{figb1}, but here we consider the model including both cosmic weak lensing and gravitation waves $r=0.1$ for the numerical simulations.}
\label{figb3}
\end{figure}

\begin{table}[!htb]
\centering
\begin{tabular}{c @{\extracolsep{0.3em}} |c c c c c c}
\hline\hline
 &  $\theta_s=10'$  & $\theta_s=20'$  &  $\theta_s=30'$  &  $\theta_s=40'$ &  $\theta_s=50'$  &  $\theta_s=60'$\\
\hline
$r=0$  & $~~0.0000\pm0.0047$ &  $~~0.0000\pm0.0060$ &  $~~0.0000\pm0.0071$ & $~~0.0001\pm0.0083$  & $~~0.0001\pm0.0094$  &  $~~0.0000\pm0.0107$ \\
  & $~~0.0000\pm0.0025$ &  $-0.0002\pm0.0039$ &  $-0.0003\pm0.0054$ & $-0.0001\pm0.0069$  & $-0.0001\pm0.0084$  &  $-0.0003\pm0.0101$ \\
\hline
$r=0.01$ & $-0.0002\pm0.0046$ &  $-0.0003\pm0.0059$ &  $-0.0002\pm0.0047$ & $-0.0001\pm0.0097$  & $-0.0004\pm0.0129$  &  $~~0.0000\pm0.0171$ \\
  & $~~0.0000\pm0.0025$ &  $-0.0001\pm0.0041$ &  $-0.0001\pm0.0061$ & $~~0.0000\pm0.0085$  & $~~0.0001\pm0.0119$  &  $~~0.0004\pm0.0160$ \\
\hline
$r=0.05$ & $-0.0002\pm0.0053$ &  $-0.0004\pm0.0086$ &  $-0.0004\pm0.0149$ & $-0.0004\pm0.0229$  & $-0.0005\pm0.0314$  &  $-0.0007\pm0.0398$ \\
  & $~~0.0001\pm0.0033$ &  $~~0.0002\pm0.0077$ &  $~~0.0005\pm0.0143$ & $-0.0009\pm0.0223$  & $~~0.0013\pm0.0307$  &  $~~0.0018\pm0.0391$ \\
\hline
$r=0.1$ & $-0.0001\pm0.0066$ &  $-0.0004\pm0.0134$ &  $-0.0008\pm0.0226$ & $-0.0011\pm0.0316$  & $-0.0015\pm0.0401$  &  $-0.0019\pm0.0480$ \\
  & $~~0.0002\pm0.0055$ &  $~~0.0005\pm0.0133$ &  $~~0.0010\pm0.0228$ & $~~0.0015\pm0.0321$  & $~~0.0019\pm0.0409$  &  $~~0.0023\pm0.0492$ \\
\hline
\end{tabular}
\caption{The skewness values (mean value and the standard deviation) of for various $B$-mode polarization maps. For each case, the upper one shows the results derived from 1000 simulated lensed $B$-map, while the lower one shows those derived from 1000 simulated Gaussian $B$-maps.}
\label{tableb1}
\end{table}

\begin{table}[!htb]
\centering
\begin{tabular}{c @{\extracolsep{0.3em}} |c c c c c c}
\hline\hline
 &  $\theta_s=10'$  & $\theta_s=20'$  &  $\theta_s=30'$  &  $\theta_s=40'$ &  $\theta_s=50'$  &  $\theta_s=60'$\\
\hline
$r=0$  & $~~1.0228\pm0.0161$ &  $~~0.5377\pm0.0159$ &  $~~0.3031\pm0.0157$ & $~~0.1834\pm0.0162$  & $~~0.1164\pm0.0176$  &  $~~0.0777\pm0.0196$ \\
  & $-0.0002\pm0.0040$ &  $~~0.0000\pm0.0067$ &  $~~0.0004\pm0.0093$ & $~~0.0006\pm0.0120$  & $~~0.0007\pm0.0149$  &  $~~0.0006\pm0.0179$ \\
\hline
$r=0.01$ & $~~0.9684\pm0.0157$ &  $~~0.4714\pm0.0154$ &  $~~0.2371\pm0.0155$ & $~~0.1240\pm0.0169$  & $~~0.0660\pm0.0196$  &  $~~0.0359\pm0.0234$ \\
  & $-0.0002\pm0.0041$ &  $~~0.0000\pm0.0070$ &  $-0.0001\pm0.0102$ & $-0.0001\pm0.0139$  & $-0.0007\pm0.0182$  &  $-0.0018\pm0.0229$ \\
\hline
$r=0.05$ & $~~0.7879\pm0.0190$ &  $~~0.2991\pm0.0177$ &  $~~0.1100\pm0.0186$ & $~~0.0394\pm0.0239$  & $~~0.0111\pm0.0317$  &  $-0.0020\pm0.0405$ \\
  & $-0.0003\pm0.0045$ &  $-0.0009\pm0.0087$ &  $-0.0022\pm0.0150$ & $-0.0045\pm0.0232$  & $-0.0074\pm0.0322$  &  $-0.0106\pm0.0414$ \\
\hline
$r=0.1$ & $~~0.6249\pm0.0231$ &  $~~0.1885\pm0.0194$ &  $~~0.0524\pm0.0232$ & $~~0.0102\pm0.0317$  & $-0.0057\pm0.0410$  &  $-0.0136\pm0.0503$ \\
  & $-0.0007\pm0.0054$ &  $-0.0023\pm0.0122$ &  $-0.0049\pm0.0219$ & $-0.0082\pm0.0324$  & $-0.0116\pm0.0424$  &  $-0.0151\pm0.0521$ \\
\hline
\end{tabular}
\caption{The kurtosis values (mean value and the standard deviation) of for various $B$-mode polarization maps. For each case, the upper one shows the results derived from 1000 simulated lensed $B$-map, while the lower one shows those derived from 1000 simulated Gaussian $B$-maps.}
\label{tableb2}
\end{table}

\section{$E$- and $B$-mode separation on an incomplete sky \label{appendixA}}

In real CMB observations, polarization components include the $Q(\hat{\gamma})$ and $U(\hat{\gamma})$. In the full-sky case, the translation from $Q$ and $U$-maps to the scalar fields $E(\hat{\gamma})$ and $B(\hat{\gamma})$ can be carried out by the standard way introduced in Sec. II.  From the definition, we know that we need the full-sky information of $Q$ and $U$ to construct the $E$- and $B$-mode fields. Unfortunately, in practice, we cannot work with full sky polarization maps due to foreground contaminations. It is always necessary to mask these contaminated regions to reduce the influence of the foregrounds. So, the mixture of $E$- and $B$-mode might become an important obstacle for the $B$-mode detection \cite{ebmixture1}.

In order to separate the $E$- and $B$-mode on the partial sky, various practical methods have been suggested in the literature \cite{ebmixture2,ebmixture3}. In this Appendix, we shall consider the method suggested by Smith and Zaldarriaga (i.e. the SZ method) \cite{ebmixture3} as an example to solve the $E$-$B$ mixture problem. The SZ method have used the so-called $\chi$-field framework, and can be effectively applied to the real data analysis or numerical simulations. In this method, the electric-type ($\mathcal{E}$) and magnetic-type ($\mathcal{B}$) scalar fields are defined as following
\begin{equation}\label{aa-1}
\mathcal{E}(\hat{\gamma})=-\frac{1}{2}[\bar{\eth}\bar{\eth} P_{+}(\hat{\gamma})+\eth\eth P_{+}(\hat{\gamma})],~~
\mathcal{B}(\hat{\gamma})=-\frac{1}{2i}[\bar{\eth}\bar{\eth} P_{+}(\hat{\gamma})-\eth\eth P_{+}(\hat{\gamma})],
\end{equation}
where $P_{\pm}(\hat{\gamma})=Q(\hat{\gamma})\pm U(\hat{\gamma})$. $\bar{\eth}$ and $\eth$ are the spin lowering and raising operators, respectively. To decompose $\mathcal{E}$ and $\mathcal{B}$ fields over the scalar spherical harmonics, we procedure as following,
\begin{equation}
\mathcal{E}(\hat{\gamma})=\sum_{\ell m} \mathcal{E}_{\ell m} Y_{\ell m}(\hat{\gamma}),~~\mathcal{B}(\hat{\gamma})=\sum_{\ell m} \mathcal{B}_{\ell m} Y_{\ell m}(\hat{\gamma}).
\end{equation}
These coefficients relate to the general coefficients $E_{\ell m}$ and $B_{\ell m}$ by the formulae $\mathcal{E}_{\ell m}=N_{\ell,2}E_{\ell m}$ and $\mathcal{B}_{\ell m}=N_{\ell,2}B_{\ell m}$, where $N_{\ell,s}=\sqrt{(\ell+s)!/(\ell-s)!}$.

In real observations, we must mask out a fractional portion of sky due to the foreground contamination. When considering the mask $W(\hat{\gamma})$, the pure magnetic-type polarization becomes $\mathcal{B}(\hat{\gamma})W(\hat{\gamma})$. In the SZ method, this components is constructed by the following practical way:
\begin{equation}
\mathcal{B}(\hat{\gamma})W(\hat{\gamma})=\sum_{\ell m} \mathcal{B}_{\ell m}^{pure}Y_{\ell m}^*(\hat{\gamma}),
\end{equation}
where
\begin{equation}
\mathcal{B}_{\ell m}^{pure}=-\frac{1}{2i}\int d\hat{\gamma}\left\{ P_{+}(\hat{\gamma})[\bar{\eth}\bar{\eth}(W(\hat{\gamma})Y_{\ell m}(\hat{\gamma}))]^*-P_{-}(\hat{\gamma})[{\eth}{\eth}(W(\hat{\gamma})Y_{\ell m}(\hat{\gamma}))]^*
\right\}.
\end{equation}

In principle, in all the observed region, one can completely construct the magnetic-type polarization without mixture residual or information loss. However, in the numerical calculation, one always has to slightly smooth the top-hat mask to reduce the numerical error. Here, we will prove that, by applying the SZ method, even for a slightly smoothed mask, i.e. with a small information loss, the mixture residual becomes very small comparing with the $B$-mode signal.  Here, the top-hat window function is smoothed by adopting a Gaussian kernel as suggested in \cite{kim}
\begin{equation}\label{gauss}
W_i=\left \{
\begin{aligned}
&\frac{1}{2}+\frac{1}{2}{\rm erf}\bigg(\frac{\delta_i-\frac{\delta_c}{2}}{\sqrt{2}\sigma }\bigg)  &\delta_{i}<\delta_{c}\\
&1  &\delta_{i}>\delta_{c}\\
\end{aligned}
\right.
\end{equation}
where where $\delta_i$ is the distance from each 1-value pixel to the closest 0-valued pixel in the top-hat window function,
$\sigma$ denotes the width of the smoothing kernel, and $\delta_c$ denotes the smoothing width. Let $ \beta $ denotes the jump range at $\delta_i=\delta_c $ and $\delta_i=0 $, then:
\begin{equation}
\beta=\frac{1}{2}-\frac{1}{2}{\rm erf}\bigg(\frac{\frac{\delta_c}{2}}{\sqrt{2}\sigma }\bigg),
\end{equation}
which is a small and adjustable parameter, and set as $\beta=0.0001$ in our calculation.

In order to show the effect of the mixture leakage on the $B$-mode polarization, we adopt the cosmological model without weak lensing and set the $r=0$, i.e. without $B$-mode polarization in the input model. We simulate 1000 random samples of $Q$ and $U$ with $\rm{FWHM}=30'$, and adopt the smoothed M1 mask. By applying the SZ method, we calculate the map $\mathcal{B}(\hat{\gamma})W(\hat{\gamma})$, which is totally caused by the leakage residuals. In Fig. \ref{figa1}, we plot the averaged power spectra $D_{\ell}\equiv \frac{1}{2\ell+1}\sum_{m}\langle |\mathcal{B}_{\ell m}^{pure}|^2\rangle$ of the maps $\mathcal{B}(\hat{\gamma})W(\hat{\gamma})$, where we have considered the values of $\delta_c=0.2^{\circ}$, $0.5^{\circ}$, $1^{\circ}$ and $1.5^{\circ}$, respectively. For the comparison, we repeat the previous calculation but now including the effects of cosmic weak lensing and gravitational waves with $r=0.1$ in the input model. The corresponding power spectra $D_{\ell}$ are also presented in Fig. \ref{figa1}. This figure clearly shows that, if the $\delta_c$ is larger, the leakage is smaller. Even for the most conservative smoothing case with $\delta_c=0.2^{\circ}$, we also find that nearly in all the multipole range, the amplitudes of the leakage are more than two orders of magnitude smaller than those of the signal. 

\begin{figure}[t]
\centerline{\includegraphics[width=12cm]{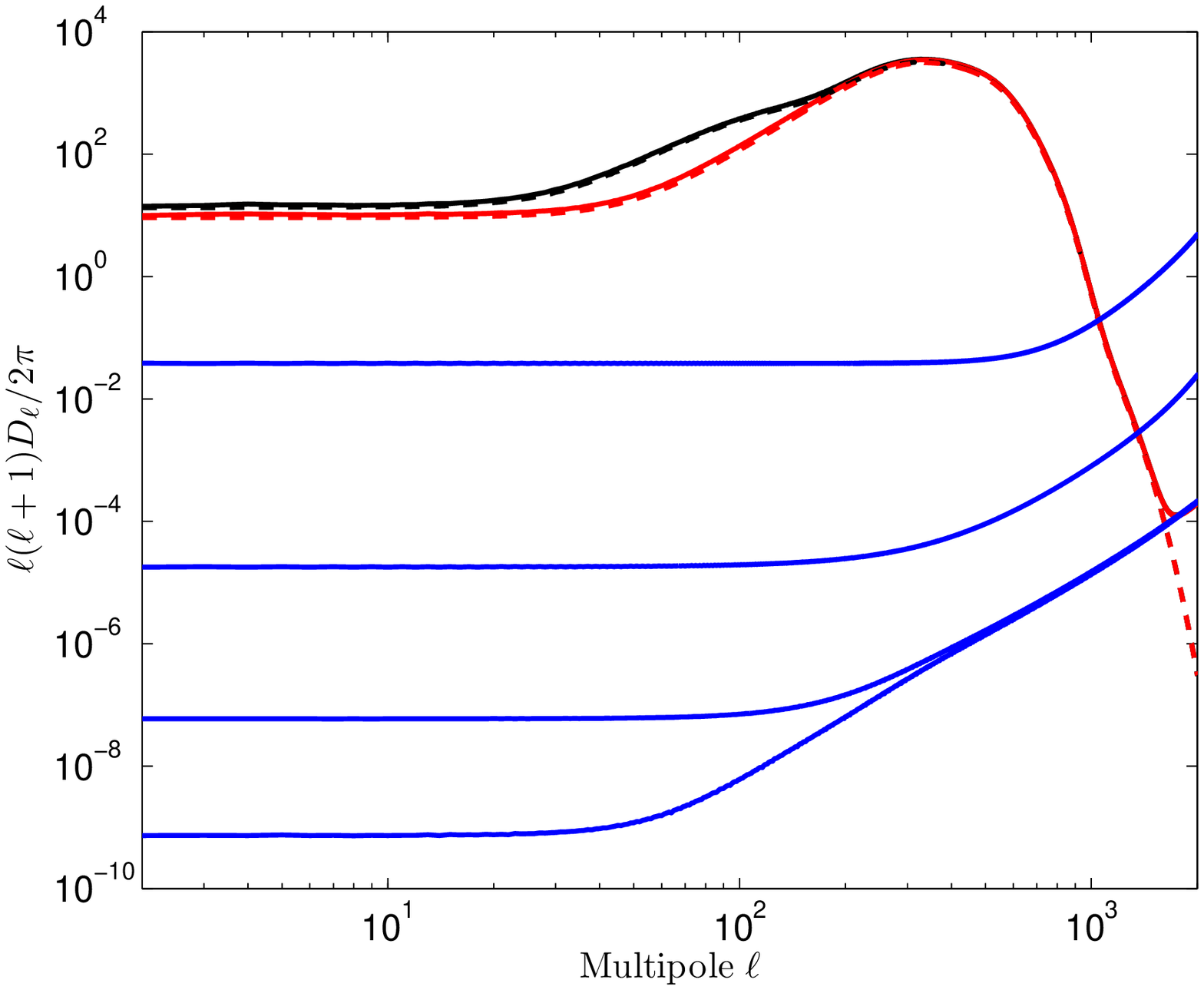}}
\caption{The averaged power spectra of the 1000 constructed maps $\mathcal{B}W$. The blue curves show that results derived from the input model without weak lensing or gravitational waves, so they are totally caused by $E$-$B$ mixture leakage. From upper to lower, the curves present the results with smoothing window function with $\delta_c=0.2^{\circ}$, $0.5^{\circ}$, $1^{\circ}$ and $1.5^{\circ}$, respectively. The black solid line denotes the spectra derived from the case with input model including weak lensing and gravitational waves $r=0.1$, and the red solid line denotes the spectra derived from the case with input model including weak lensing only. In both cases, we have set $\delta_c=1^{\circ}$ for the smoothing window functions. Note that, the black and red dashed lines are the analytical power spectra for these two cases, respectively. }
\label{figa1}
\end{figure}

To study the effects of the $E$-$B$ leakage on the $B$-mode statistical properties, we compare the MFs in the following two cases, i.e. ideal case and real case. In the ideal case, for a given cosmological model with $r=0$, we simulate 1000 lensed CMB samples. Each sample includes a pair of full-sky $Q$-map and $U$-map. By using the formulae in Eq. (\ref{aa-1}), we construct the full-sky $\mathcal{E}$- and $\mathcal{B}$-map from $Q$- and $U$-map, and mask them by applying the smoothed mask $W(\hat{\gamma})$ with parameters $\beta=0.0001$, $\delta_c=1^{\circ}$. Thus, we calculate the MFs of these masked $\mathcal{B}W$-maps, which are denoted as $v_i^{\rm ideal}(\nu)$ with $i=0,1,2$. While, in the real case, we do the following procedure: For each sample, we use the simulated full-sky $Q$- and $U$-maps derived from the ideal case, and mask them with smoothed mask $W(\hat{\gamma})$. These masked maps mimic the potential observations. Then, by the SZ method we construct the quantities $\mathcal{B}_{\ell m}^{pure}$ and the corresponding $\mathcal{B}W$-map. We calculate the MFs of these masked $\mathcal{B}W$-maps, which are denoted as $v_i^{\rm real}(\nu)$. The difference between these two $\mathcal{B}W$-maps is that the $\mathcal{B}W$-map in the real case includes the $E$-$B$ leakage, while the map in the ideal case is free from it. In order to reduce the $E$-$B$ leakage, in both cases we simulate the maps with resolution parameter $N_{\rm side}=1024$, maximum multipole $\ell_{\max}=2048$ and ${\rm FWHM}=30'$.
\begin{figure}[t]
\centerline{\includegraphics[width=14cm]{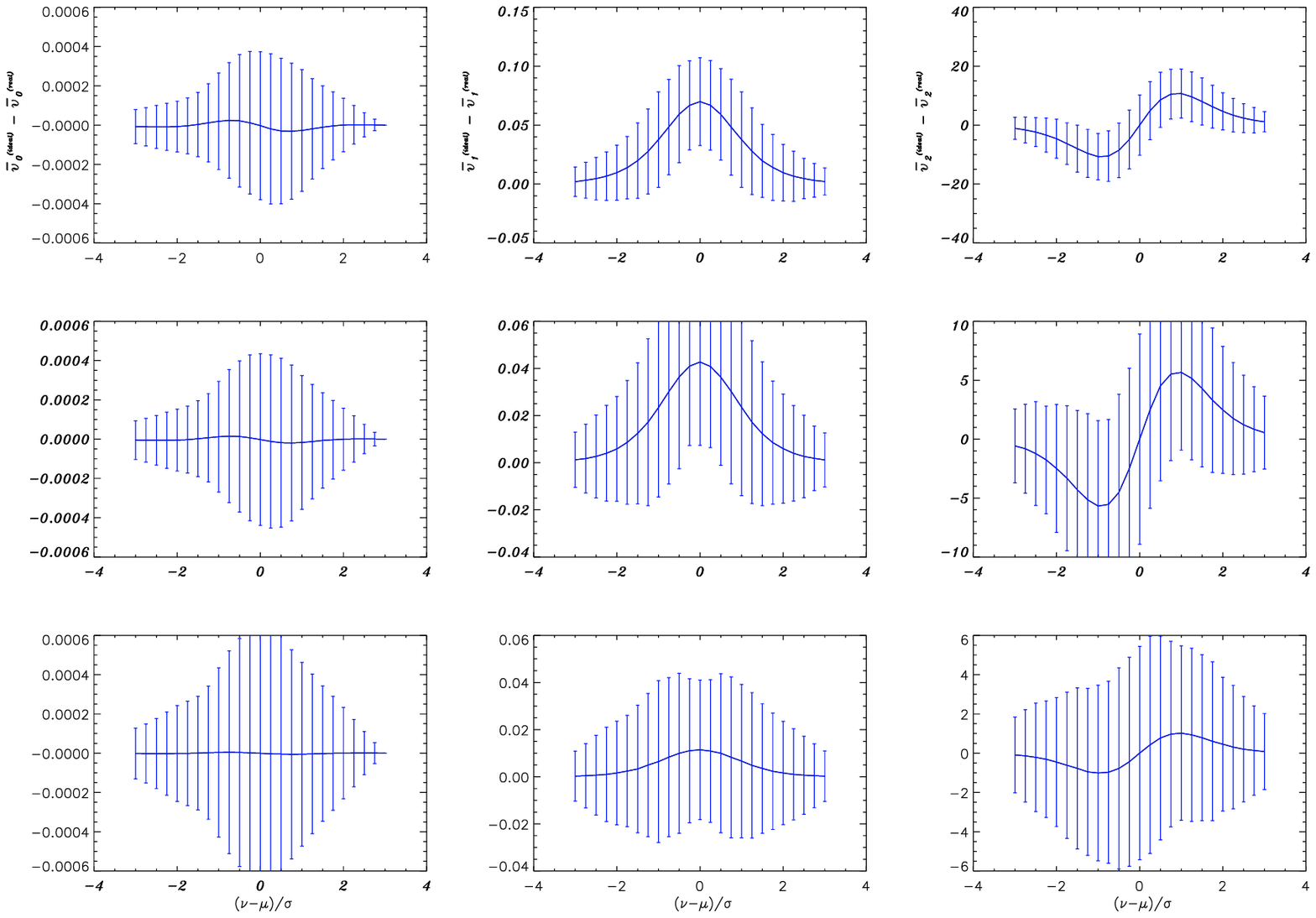}}
\caption{The difference between the MFs $v_i$ of in ideal case (without $E$-$B$ leakage) and real case (with $E$-$B$ leakage). The upper panels show the results with smoothing scale $\theta_s=10'$, and middle panels show those with $\theta_s=40'$ and lower panels show those with $\theta_s=60'$. For all the panels, we have considered the cosmological model with $r=0$, and applied the smoothed M1 mask (see main text for the details).}
\label{figa2}
\end{figure}

In Fig. \ref{figa2}, we plot the differences $\bar{v}_i^{\rm ideal}-\bar{v}_i^{\rm real}$ for the case with different smoothing scale $\theta_s$, which shows the influence of leakage on the MFs. We find that for each $\theta_s$, the influence caused by the leakage is quite small compared with the intrinsic non-Gaussianity in the lensed $B$-map. In order to quantify it, we define the following quantity
 \begin{equation}
 \hat{\chi}_{\theta_s}^2=\sum_{\alpha \alpha'}[\bar{v}_{\alpha}^{\rm ideal}-\bar{v}_{\alpha}^{\rm real}]\Sigma_{\alpha\alpha'}^{-1}[\bar{v}_{\alpha'}^{\rm ideal}-\bar{v}_{\alpha'}^{\rm real}],
 \end{equation}
where $\alpha$ and $\alpha'$ denote the binning number of the threshold value $\nu$ and the different kinds of MF $i$, and $\Sigma$ is the corresponding covariance matrix. For the case with $\theta_s=10'$, we get $\hat{\chi}_{\theta_s}^2=17.8$, while for $\theta_s=40'$, we have $\hat{\chi}_{\theta_s}^2=1.6$, and for $\theta_s=60'$, we have $\hat{\chi}_{\theta_s}^2=0.3$. Comparing with the intrinsic non-Gaussianity of the lensed $B$-maps (see the results in Table \ref{table1} and Fig. \ref{fig5}), we find the influence caused by the $E$-$B$ leakage is quite small.

In the end of this Appendix, we should mention that the $E$-$B$ leakage might be quite different for different $E$-$B$ separation methods \cite{ebmixture2,ebmixture3}. Here, we have adopted the SZ method, which can only construct pure $\mathcal{B}$-map, instead of the general $B$-map. However, it would be possible to construct the pure $B$-map for some other methods (for instance, the method suggested by Bunn et al. in \cite{ebmixture2}). In a separate paper \cite{larissa}, we will detailed discuss the influence of leakage for different $E$-$B$ separating methods, different cosmological models, and different noise levels. Meanwhile, we will also discuss the possible way to reduce the leakage with the price of more information loss. However, these are outside the scope of this paper.

\section{Constructing the estimators of CMB power spectrum based on the incomplete maps  \label{appendixC}}

 A given random two-dimensional scalar field on the sphere $S(\hat{\gamma})$ (it might be $T$-map, $E$-map or $B$-map) can be decomposed as the standard spherical harmonics as follows,
\begin{equation}
S(\hat{\gamma})=\sum_{\ell m} s_{\ell m} Y_{\ell m} (\hat{\gamma}),
\end{equation}
where $Y_{\ell m}(\hat{\gamma})$ are the spherical harmonics, and $s_{\ell m}$ are the corresponding coefficients. Then, the power spectrum $C_{\ell}$ is defined as
\begin{equation}
C_{\ell}\equiv \frac{1}{2\ell+1} \sum_{m=-\ell}^{\ell} \langle s_{\ell m} s^*_{\ell m}\rangle,
\end{equation}
where $\langle ... \rangle$ denotes the average over the statistical ensemble of realizations. From the observable map $S(\hat{\gamma})$, one can define the estimator $\hat{C}_{\ell}=\frac{1}{2\ell+1}\sum_{m}s_{\ell m} s^*_{\ell m}$, which is the best-unbiased estimator for the power spectrum $C_{\ell}$ \cite{grishchuk1997}.

However, if considering the mask window function $W(\hat{\gamma})$ applied to the original full sky map $S(\hat{\gamma})$, the construction of the unbiased estimator for $C_{\ell}$ is not straightforward. A large number of methods have been suggested in the literature \cite{method1,method2,method3,method4,method5}. In this Appendix, we adopt the so-called pseudo-$C_{\ell}$ (PCL) estimator method \cite{method4}. Although the PCL estimator is a suboptimal one, it can be easily applied in pixel space using fast spherical harmonics transformation, and it has been used in various CMB observations including in both WMAP and Planck data.

Considering the window function $W(\hat{\gamma})$, the pseudo coefficients $\tilde{s}_{\ell m}$ can be defined as
\begin{equation}
\tilde{s}_{\ell m}=\int S(\hat{\gamma}) W(\hat{\gamma}) Y_{\ell m}(\hat{\gamma})d\hat{\gamma},
\end{equation}
which is related to $s_{\ell m}$ by
\begin{equation}
\tilde{s}_{\ell m}= \sum_{\ell_{1}m_{1}} s_{\ell_{1}m_{1}} K_{\ell m\ell_{1}m_{1}}.
\end{equation}
The coupling matrix $K$ is given by
\begin{equation}
K_{\ell m\ell_{1}m_{1}}=\sqrt{\frac{(2\ell_1+1)(2l+1)}{4\pi}}\sum_{\ell_2 m_2} (-1)^{m}(2\ell_2+1) w_{\ell_{2}m_{2}}
\begin{pmatrix}
\ell_1 & \ell_2 & \ell \\
0 & 0 & 0
\end{pmatrix}
\begin{pmatrix}
\ell_1 & \ell_2 & \ell \\
m_1 & m_2 & -m
\end{pmatrix},
\end{equation}
and $w_{\ell m}$ are the coefficients of spherical harmonics expansion of the mask $W(\hat{\gamma})$, i.e.,
\begin{equation}
w_{\ell m}=\int W(\hat{\gamma}) Y_{\ell m}^* (\hat{\gamma}) d\hat{\gamma}.
\end{equation}

The pseudo estimator $\tilde{C}_{\ell}$ is defined in terms of the multipole coefficients $\tilde{s}_{\ell m}$ as
\begin{equation}
\tilde{C}_{\ell}=\frac{1}{2\ell+1} \sum_{m=-\ell}^{\ell} \tilde{s}_{\ell m} \tilde{s}^*_{\ell m}.
\end{equation}
The expectation value of $\tilde{C}_{\ell}$ is
$\langle \tilde{C}_{\ell} \rangle =\sum_{\ell'}C_{\ell'} M_{\ell \ell'}$,
where the coupling matrix is
\begin{equation}
M_{\ell \ell'}=(2\ell'+1)\sum_{\ell_{2}} \frac {2\ell_{2}+1} {4\pi}
\begin{pmatrix}
    \ell'&\ell_2&\ell \\
    0 & 0& 0
    \end{pmatrix}  ^{2}\tilde{w}_{\ell_{2}}
\end{equation}
and ${\tilde w}_{\ell}$ are the following power spectrum,
\begin{equation}
\tilde{w}_{\ell} = \frac{1}{2\ell+1}\sum_{m=-\ell}^{\ell}w_{\ell m}w^{*}_{\ell m}.
\end{equation}
So, the unbiased estimator for spectrum $C_{\ell}$ in the masked sky can be constructed as
$\hat{C}_{\ell}=\sum_{\ell'} M^{-1}_{\ell \ell'} \tilde{C}_{\ell'}$.

\end{document}